\documentclass[12pt]{article}

\usepackage{scicite}
\usepackage{times}
\usepackage{amssymb}
\usepackage{graphicx}
\usepackage{dcolumn}
\usepackage{bm}
\usepackage{amsmath}
\usepackage{mathrsfs}
\usepackage{textcomp}

\def\be{\begin{equation}}
\def\ee{\end{equation}}
\def\bea{\begin{eqnarray}}
\def\eea{\end{eqnarray}}

\newenvironment{sciabstract}{%
\begin{quote} \bf}
{\end{quote}}

\title{Study to improve the performance of interferometer with ultra-cold atoms}

\author
{ Xiangyu Dong$^{1}$, Shengjie Jin$^{1}$, Hongmian Shui$^{1}$, Peng Peng$^{1}$, \\
Xiaoji Zhou$^{1,\star}$\\
\normalsize{$^{1}$State Key Laboratory of Advanced Optical Communication System and Network,} \\
\normalsize{Department of Electronics, Peking University, Beijing 100871, China}
}
\date{\today}

\begin{document}


\baselineskip24pt


\maketitle

\begin{sciabstract}
Ultra-cold atoms provide ideal platforms for interferometry. The macroscopic matter-wave property of ultra-cold atoms leads to large coherent length and long coherent time, which enable high accuracy and sensitivity to measurement.
Here, we review our efforts to improve the performance of the interferometer. We demonstrate a shortcut method for manipulating ultra-cold atoms in an optical lattice. Compared with traditional ones, this shortcut method can reduce manipulation time by up to three orders of magnitude. We construct a matter-wave Ramsey interferometer for trapped motional quantum states and significantly increase its coherence time by one order of magnitude with an echo technique based on this method.
Efforts have also been made to enhance the resolution by multimode scheme.
Application of a noise-resilient multi-component interferometer shows that increasing the number of paths could sharpen the peaks in the time-domain interference fringes, which leads to a resolution nearly twice compared with that of a conventional double-path two-mode interferometer. With the shortcut method mentioned above, improvement of the momentum resolution could also be fulfilled, which leads to atomic momentum patterns less than 0.6 $\hbar k_L$.
To identify and remove systematic noises, we introduce the methods based on the principal component analysis (PCA) that reduce the noise in detection close to the $1/\sqrt{2}$ of the photon-shot noise and separate and identify or even eliminate noises.
Furthermore, we give a proposal to measure precisely the local gravity acceleration within a few centimeters based on our study of ultracold atoms in precision measurements.
\end{sciabstract}


\section{Introduction}
Precision measurement is the cornerstone of the development of modern physics. Atom-based precision measurement is an important part.
In recent years, ultra-cold atoms have attracted extensive interest in different fields~\cite{XuRN32,CooperRN203,RN114,MazurenkoRN164}, because of their macroscopic matter-wave property~\cite{1dalfovo1999theory,BlochRN78}. This property will lead to large coherent length and long coherent time, which enable high fringe contrast~\cite{2005.00368,Hardman2016,Hardman2014}.
Hence, with these advantages, ultra-cold atoms provide ideal stages for precision measurement~\cite{Ye2007,RN257} and have numerous applications~\cite{SchrepplerRN228,XuRN32,SchrepplerRN228,Xiong_2013,Hardman2016,ColloquiumRN80,Campbell2017,PhysRevA.81.012115}, ranging from inertia measurements~\cite{Hardman2016,XuRN32,SchrepplerRN228} to precision time keeping~\cite{ColloquiumRN80,Campbell2017}.
For example, a Bose-Einstein condensate is used as an atomic source for a high precision sensor, which is released into free fall for up to 750 ms and probed with a $T=130$ ms Mach-Zehnder atom interferometer based on Bragg transitions~\cite{Hardman2016}.
A trapped geometry is realized to probe gravity by holding ultra-cold cesium atoms for 20 seconds~\cite{XuRN32}, which suppresses the phase variance due to vibrations by three to four orders of magnitude, overcoming the dominant noise source in atom-interferometric gravimeters.
With ultra-cold atoms in an optical cavity, the detection of weak force can achieve a sensitivity of 42 $\rm{yN}/\sqrt{\rm{Hz}}$, which is a factor of 4 above the standard quantum limit~\cite{SchrepplerRN228}.

With the superiority mentioned above, it is obvious to consider an interferometer by ultra-cold atoms to further precision measurements. Interferometers with atoms propagating in free fall are ideally suited for inertia measurements~\cite{PhysRevLett.124.120403,PhysRevA.80.063604,LeRN260,SchmidtRN261,PhysRevLett.107.133001,TackmannRN263,AltinRN262}. Meanwhile, with atoms held in tight traps or guides, they are better to measure weak localized interactions. For example, a direct measurement to the Casimir-Polder force is performed by I. Carusotto et al. in 2005, which is as large as $10^{-4}$ gravity~\cite{PhysRevLett.95.093202}. However, ultra-cold atoms still get some imperfections needed to be surmounted when combining with interferometry. The macroscopic matter-wave property is generated simultaneously with non-linear atom-atom interactions. Phase diffusion caused by interactions limits the coherence time, and ultimately restricts the sensitivity of interferometers. Besides the interrogation time, the momentum splitting as well as the path number also has an impact on the sensitivity. It has been demonstrated by experiments of multipath interferometers that, interferometric fringes can be sharpened due to the higher-harmonic phase contributions of the multiple energetically equidistant Zeeman states~\cite{7.4weitz1996multiple,10petrovic2013multi}, whereas a decrease in the average number of atoms per path causes a greater susceptibility to shot noise. Equilibrium between these parameters could lead to an optimal resolution. In addition, we should also pay attention to the signal analysis procedure as the interferometric information is mainly extracted from the signal detected. The resulting resolution severely relies on the probing system.

In this review, we mainly introduce our experimental developments that study these fundamental and important
issues to improve the performance of interferometer with ultra-cold atoms. The main developments are concentrated in three aspects: increasing coherent time, using multimode scheme and reducing systematic noises.

\textbf{A. Enhanced resolution by increasing coherent time.} We introduce an effective and fast (few microseconds) methods, for manipulating ultra-cold atoms in an optical lattice (OL), which can be used to construct the atomic interferometer and increase the coherent time to finally get a higher resolution. This shortcut loading method is a designed pulse sequence, which can be used for preparing and manipulating arbitrary pure states and superposition states. Another advantage of this method is that the manipulation time is much shorter than traditional methods (100 ms$\to$100 $\rm{\mu s}$).
Based on this shortcut method, we constructed an echo-Ramsey interferometer (RI) with motional Bloch states (at zero quasi-momentum on S- and D-bands of an OL)~\cite{Hu2018}. Thanks to the rapidity of shortcut methods, more time could be used for the RI process. We identified the mechanisms that reduced the RI contrast, and greatly increased the coherent time (1.3 ms $\to$14.5 ms) by a quantum echo process, which eliminated the influence of contrast attenuation mechanisms mostly.

\textbf{B. Enhanced resolution by multimode scheme.} Several efforts have been made to avoiding the decays of interferometric resolution because of the experimental noises. We demonstrated that the improvement of the phase resolution could be accomplished by a noise-resilient multi-component interferometric scheme. With the relative phase of different components remaining stable, increasing the number of paths could sharpen the peaks in the interference fringes, which led to a resolution nearly twice compared with that of a conventional double-path two-mode interferometer. Moreover, improvement of the momentum resolution was fulfilled with
optical lattice pulses. We got results of atomic momentum patterns with intervals less than the double recoil momentum. The momentum pattern exhibited 10 main peaks.

\textbf{C. Enhanced resolution by removing the systematic noise.} The method to identify and remove systematic noises for ultra-cold atoms is also introduced in this paper. For improving the quality of absorption image, which is the basic detection result in ultra-cold atoms experiments, we developed an optimized fringe removal algorithm (OFRA), making the noise close to the theoretical limit as $1/\sqrt{2}$ of the photon-shot noise.
Besides, for the absorption images after preprocessing by OFRA, we applied the principal component analysis to successfully separate and identify noises from different origins of leading contribution, which helped to reduce or even eliminate noises via corresponding data processing procedures.

Furthermore, based on our study of ultracold atoms in precision measurements, we demonstrated a scheme for potential compact gravimeter with ultra-cold atoms in a small displacement.

The text structure is as follows.
In Sec.2, a shortcut method manipulating ultra-cold atoms in an optical lattice and an Echo-Ramsey interferometry with motional quantum states are introduced, which can increase the coherent time. In Secs.3, we prove that the resolution can be increased using a double-path multimode interferometer with spinor Bose-Einstein condensates (BECs) or an optical pulse, both of which can be classified into multimode scheme. In Secs.4, methods for identifying and reducing the systematic noises for ultra-cold atoms are demonstrated. Finally, we give a proposal on gravimeter with ultra-cold atoms in Secs.5.

\section{\label{sec:Coherent Time}Enhanced resolution by increasing coherent time}
\begin{figure}
\begin{center}
\includegraphics[width=0.8\linewidth]{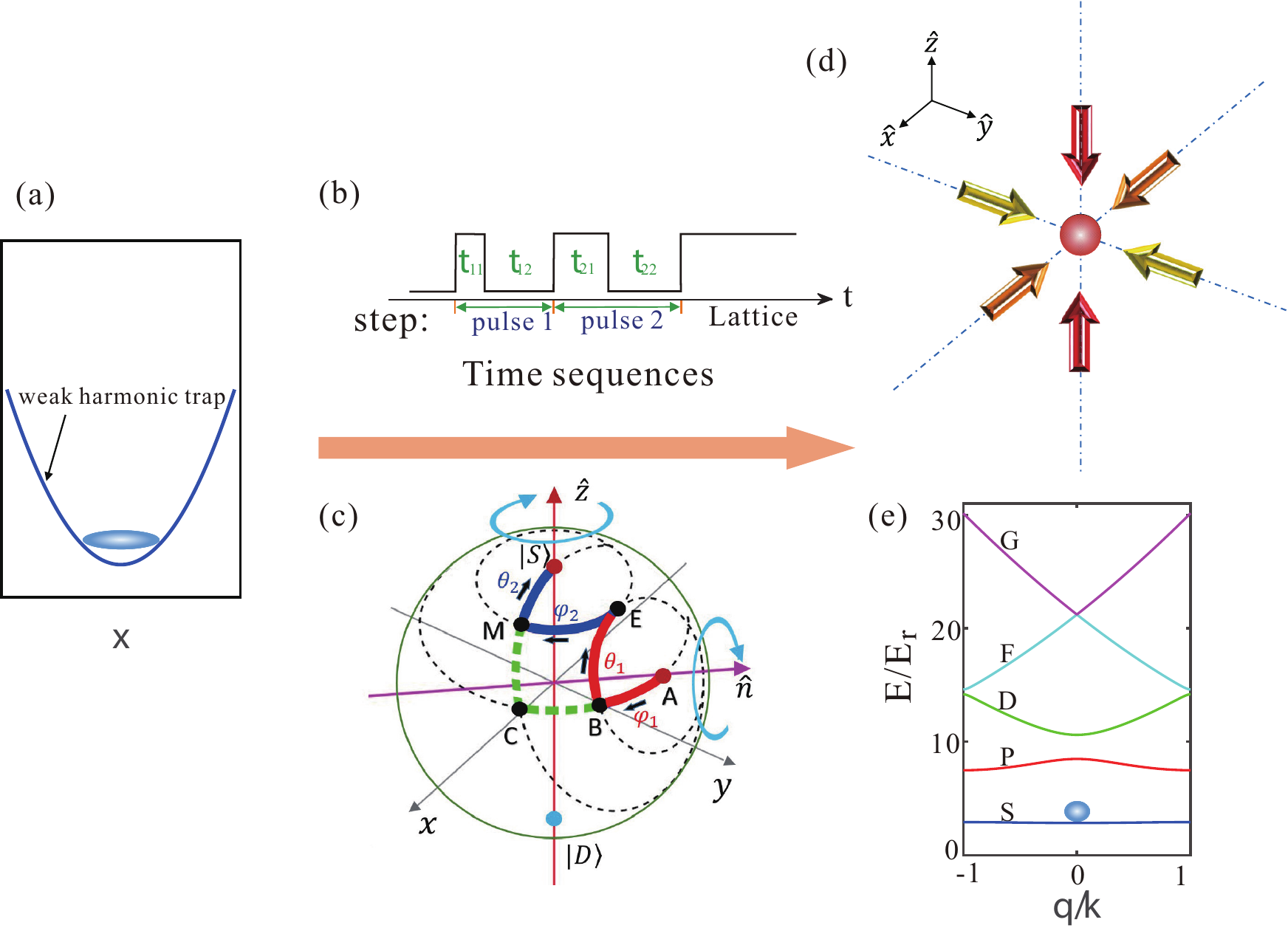}
\end{center}
\caption{Schematic diagram of the shortcut method (take ground state preparation as an example).
(a) At the beginning, the BECs are formed in a weak harmonic trap.
(b) Time sequence of shortcut method.
(c) Mapping the shortcut process onto the Bloch sphere. The track $A\to C \to | S \rangle$ and track $A\to B \to E \to M \to | S \rangle$ represent one pulse and two pulses shortcut process, respectively.
(d) After this shortcut process, the desired states of an 3D optical lattice are prepared.
(e) Band structure of 1D OL with different quasi-momentum $q$ when $V_0=10\;E_r$.
Reproduced with permission from Ref.~\cite{Zhou2018}.
}
\label{fig:shortcut1}
\end{figure}

The macroscopic coherent properties of ultra-cold atoms~\cite{PhysRevA.83.051608,Wang13,PhysRevA.83.053603,PhysRevA.81.013615,Zhou10,PhysRevA.80.063608,PhysRevA.79.033605,LI20084750,PhysRevA.78.052107,PhysRevA.78.043611,PhysRevA.83.033620} are conducive to precise measurement. To make full use of the advantages of ultra-cold atomic coherence properties, one method is to reduce the manipulation time and another is to suppress the attenuation of coherence.

Firstly, we demonstrated a shortcut process for manipulating BECs trapped in an OL~\cite{Zhou2018,FF, Chen}. By optimizing the parameters of the pulses, which constitute the sequence of the shortcut process, We can get extremely high fidelity and robustness for manipulating BECs into the desired states, including the ground state, excited states, and superposition states of a one, two or three-dimensional OLs. Another advantage of this method is that the manipulation time is much shorter than that in traditional methods (100 ms$\to$100 $\rm{\mu s}$).

This shortcut is composed of optical lattice pulses and intervals that are imposed on the system before the lattice is switched on. The time durations and intervals in the sequence are optimized to transfer the initial state to the target state with high fidelity. This shortcut procedure can be completed in several tens of microseconds, which is shorter than the traditional method (usually hundreds of milliseconds). It can be applied to the fast manipulation of the superposition of Bloch states.

Then, based on this method, we constructed an echo-Ramsey interferometer (RI) with motional Bloch states (at zero quasi-momentum on S- and D-bands of an OL)~\cite{Hu2018}. The key to realizing a RI is to design effective $\pi$- and $\pi/2$ pulses, which can be obtained by the shortcut method~\cite{Liu,Zhai,Zhou2018}. Thanks to the rapidity of shortcut methods, more time can be used for the RI process. We identified the mechanisms that reduced the RI contrast, and greatly increased the coherent time (1.3 ms $\to$14.5 ms) by a quantum echo process, which eliminated the influence of most contrast attenuation mechanisms.

\subsection{Shortcut manipulating ultra-cold atoms}

Efficient and fast manipulation of BECs in OLs can be used for precise measurements, such as constructing atom-based interferometers and increasing the coherent time of these interferometers.
Here we demonstrate an effective and fast (around 100 $\rm{\mu s}$) method for manipulating BECs from an arbitrary initial state to a desired OL state.
This shortcut method is a designed pulse sequence, in which the parameters, such as duration and interval of each step, are optimized to maximize fidelity and robustness of the final state.
With this shortcut method, the pure Bloch states with even or odd parity and superposition states of OLs can be prepared and manipulated. In addition, the idea can be extended to the case of two- or three-dimensional OLs. This method has been verified by experiments many times and is very consistent with the theoretical analysis~\cite{Zhou2018,PhysRevA.88.013603,PhysRevA.83.063604,PhysRevA.88.053629,Liu,RN107,RN111,Hu2018}.

We used the simplest one-dimensional standing wave OL to demonstrate the design principle of this method. The OL potential is $V_{OL}(x)=V_0 \cos^2{kx}$, where $V_0$ is used to characterize the depth of the OL.

Supposing that the target state $\left| {\psi_a} \right\rangle$ is in the OL with depth
$V_0$, $m$-step preloading sequence has been applied on the initial state $\left|\psi_i\right\rangle$. The final states $\left| {\psi_f} \right\rangle$ is given by
\begin{equation}
\centering|\psi_f\rangle= \prod_{j=1}^{m} \hat{Q}_j|\psi_i\rangle,
\label{sequence}
\end{equation}
where $\hat{Q}_j=e^{-i\hat{H}_j t_j}$ is the evolution
operator of the $j$th step. By maximizing the fidelity
\begin{equation}
Fidelity=|\langle \psi_a|\psi_f\rangle|^2,
\label{fidelity}
\end{equation}
we can get the optimal parameters $\hat{H}_j$ and $t_j$.

This preprocess is called a shortcut method, which can be used for loading atoms into different bands of an optical lattice. For example, the shortcut loading ultra-cold atoms into S-band in a one-dimensional optical lattice is shown in Fig.~\ref{fig:shortcut1}.

By setting different initial state and target state, different time sequences can be designed to manipulate atoms, to build different interferometers, which greatly saves the coherent time. Based on this shortcut method, we can prepare exotic quantum states~\cite{RN114,RN112,Zhou2018} and construct interferometer with motional quantum states of ultra-cold atoms~\cite{Hu2018}.

\subsection{Increasing coherent time in a Ramsey interferometry with motional Bloch states of ultra-cold atoms}
Suppressing the decoherence mechanism in the atomic interferometer is beneficial for increasing coherence time and improving the measurement accuracy.
Here we demonstrated an echo method can increase the coherent time for Ramsey interferometry with motional Bloch states (at zero quasi-momentum on S- and D-bands of an OL) of ultra-cold atoms~\cite{Hu2018}. The RI can be applied to the measurement of quantum many-body effects.
The key challenge for the construction of this RI is to achieve $\pi$- and $\pi/2$-pulses, because there is no selection rule for Bloch states of OLs.
The $\pi$- or $\pi/2$-pulse sequences can be obtained by the shortcut method~\cite{FF, Chen,Liu,Zhai,Zhou2018}, which precisely and rapidly manipulates the superposition of BECs at the zero quasi-momentum on the 1st and 3rd Bloch bands. Retaining the OL, we observed the interference between states and measured the decay of coherent oscillations.

We identified the mechanisms that reduced the RI contrast: thermal fluctuations, laser intensity fluctuation, transverse expansion induced by atomic interaction, and the nonuniform OL depth. Then, we greatly increased the coherent time (1.3 ms $\to$14.5 ms) by a quantum echo process, which eliminated the influence of most contrast attenuation mechanisms.

\subsubsection{Ramsey interferometer in an optical lattice}
\begin{figure}
\begin{center}
\includegraphics[width=0.6\linewidth]{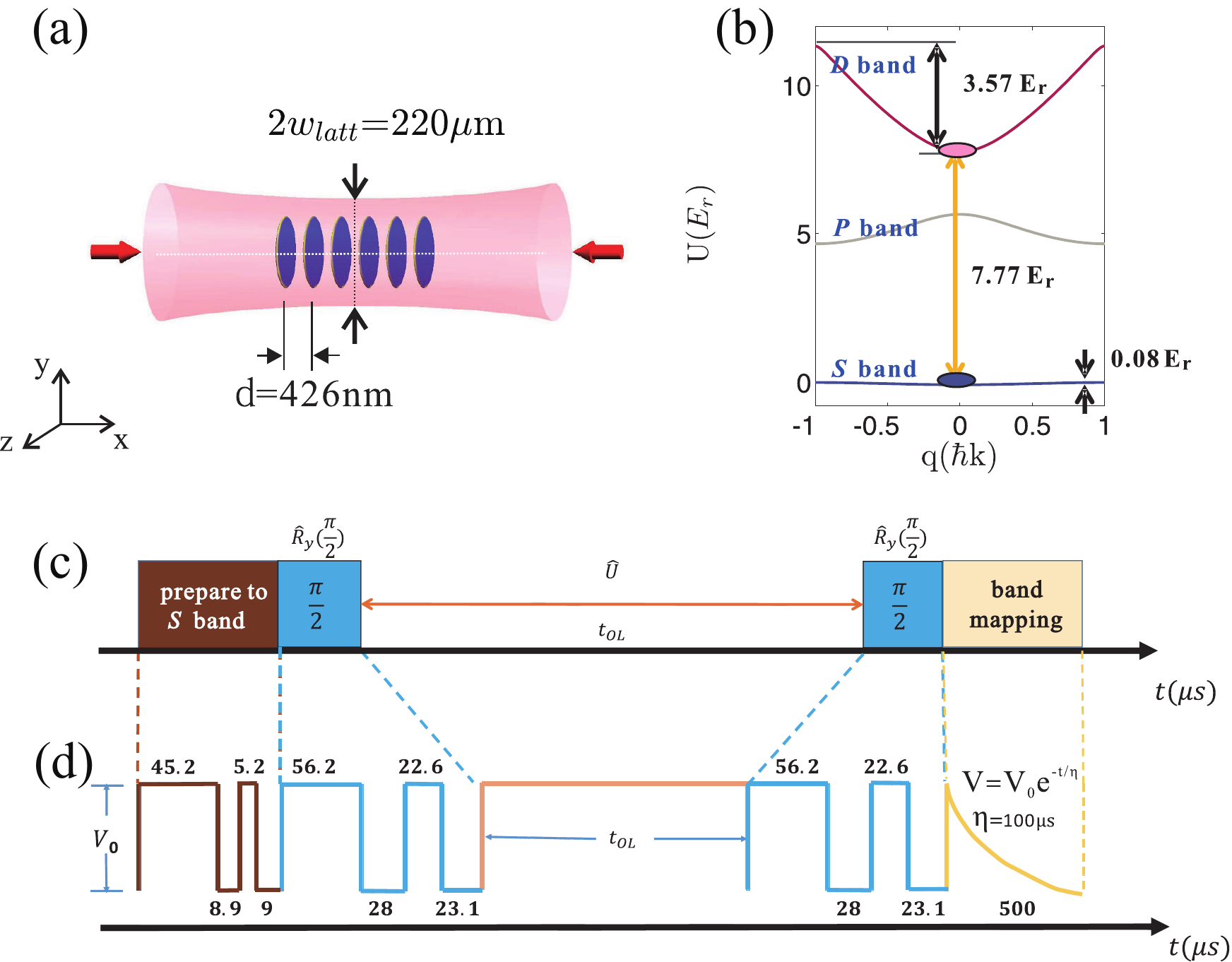}
\end{center}
\caption{Experimental configuration for a Ramsey interferometer in a $V_0=10\;E\mathrm{r}$ lattice:
(a) The BEC is divided into discrete pancakes in $yz$ plane by an 1-dimensional optical lattice along $x$ axis with a lattice constant $d=426$ nm.
(b) Band energies for the S-band and the D-band.
(c) Time sequences for the Ramsey interferometry. The atoms are first loaded into the S band of OL, followed by the RI sequence:  $\pi/2$ pulse,  holding time $t_{OL}$, and the second $\pi/2$ pulse. Finally band mapping is used to detect the atom number in the different bands.
(d) The used pulse sequences designed by an optimised shortcut method.
Reproduced with permission from Ref.~\cite{Hu2018}.}
\label{f1}
\end{figure}

This RI starts from BECs of $^{87}\rm{Rb}$ at the temperature 50 nK similar to our previous work ~\cite{NiuOE,HuPRA,GuoRN265,RN231,RN114,RN111,RN113,RN107,RN112}. Then a 1D standing wave OL is formed (Fig.~\ref{f1}). After a shortcut sequence, the BEC is transferred into the ground band of the OL, denoted as $\phi_{S,0}$.

\begin{figure*}
\begin{center}
\includegraphics[width=0.8\linewidth]{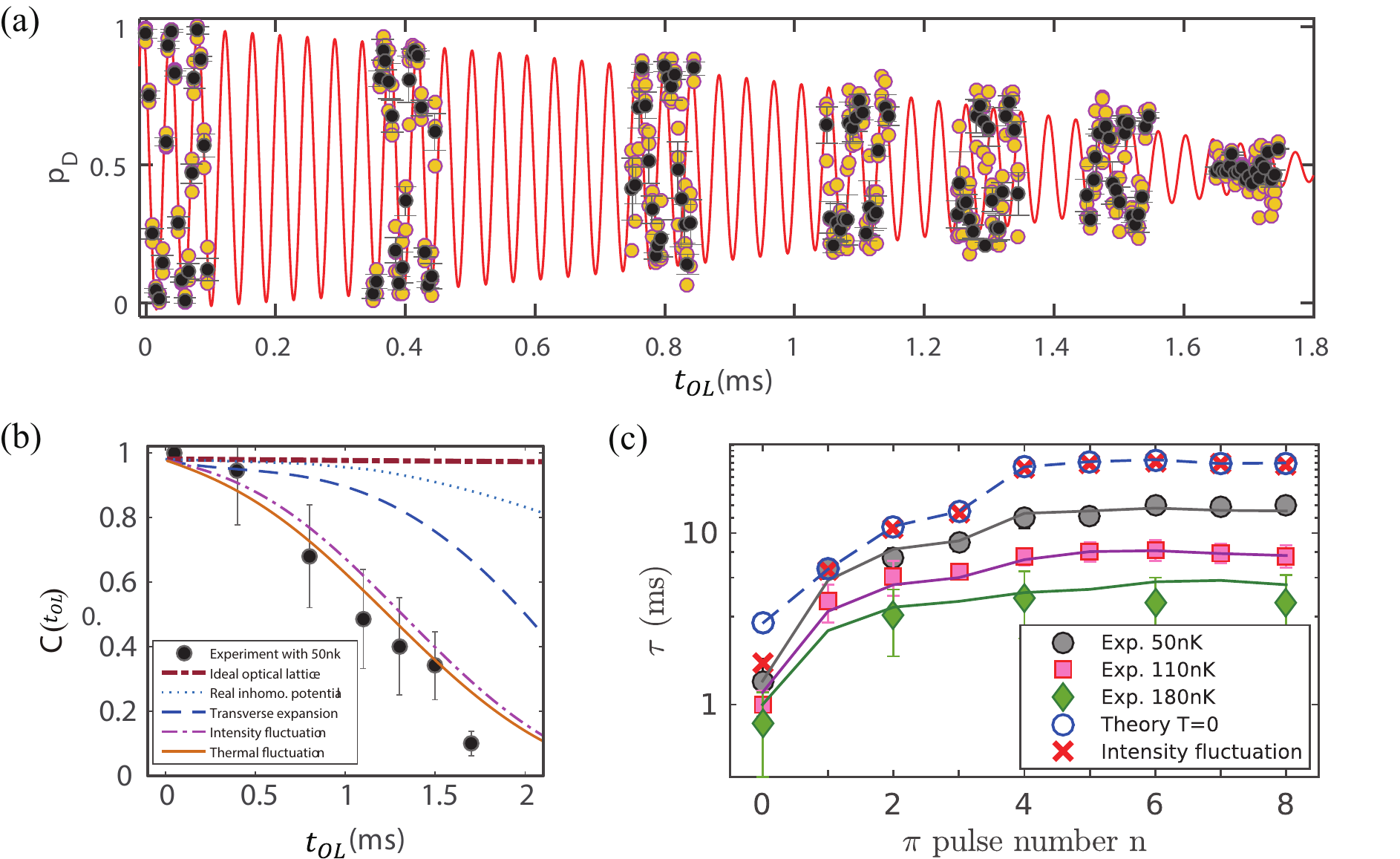}
\end{center}
\caption{
(a) Change of $p_{\mathrm{D}}$, the population of atoms in the D-band, over time $t_{OL}$ with temperature $T=50$ nK.
(b) Influence of different mechanisms on the RI.
(c) Characteristic time $\tau$ for the different number of $\pi$ pulse $n$ and different temperatures. The circles, squares, and diamonds represent the experimental results and lines are fitting curves.
Reproduced with permission from Ref.~\cite{Hu2018}.}
\label{f2}
\end{figure*}

Fig.~\ref{f1}(b) illustrates that the RI is constructed with Bloch states $\phi_{i,q}$, which includes the ground band $|S\rangle$, the third band $|D\rangle$, and their superposition state $\psi=a_S|S\rangle+a_D|D\rangle$ (denoted as $\binom{a_S}{a_D}$). It is difficult to realize the interferometer with this pseudo-spin system, because there is no selection rule for Bloch states of OLs. However, thanks to the existence of a coherent, macroscopic matter-wave, a $\pi/2$-pulse for BECs in an OL can be obtained, where the Bloch states $|S\rangle$ and $|D\rangle$ are to be manipulated to $|\psi_{1}\rangle$=$(|S\rangle+|D\rangle)/\sqrt{2}$ and $|\psi_{2}\rangle$=$(-|S\rangle+|D\rangle)/\sqrt{2}$, respectively. Fig.~\ref{f1}(d) shows the $\pi/2$ pulse we used for the RI~\cite{Liu, Zhai,Zhou2018} with fidelities of $98.5\%$ and $98.0\%$ respectively~\cite{Liu, Zhai,Zhou2018,Hu2018}.

Fig.~\ref{f1}(c) illustrates the whole process of RI. First, the BEC is transferred to $|S\rangle$ of an OL. Then, a $\pi/2$-pulse $\hat L(\pi/ 2)$ is applied to atoms, where $\hat L(\pi/2)\binom{1}{0}=\frac{1}{\sqrt{2}}\binom{1}{1}$. After holding time $t_{OL}$ and another $\pi/2$-pulse, the state is:
\begin{equation}\label{e1}
    \psi_f=\hat L(\pi/2) \hat Q(t_{OL}) \hat L(\pi/2)\psi_i,
\end{equation}
where $\hat L(\alpha)=(\cos\frac{\alpha}{2}-i\sin\frac{\alpha}{2}) \hat \sigma_y$. The operator $\hat Q(t_{OL})=(\cos\omega t+i\sin\omega t)\hat \sigma_z$ ($\omega$ corresponds to the energy gap between S and D bands at zeros quasi-momentum).

Fig.~\ref{f2}(a) depicts the results $p_{D}(t_{OL})=N_{D}/(N_{S}+N_{D})$ at different $t_{OL}$ for the RI process ($\hat R(\pi/2)-\hat U(t_{OL})-\hat R(\pi/2)$). $N_\mathrm{S}$ ($N_\mathrm{D}$) represents the number of atoms in S-band (D-band). The period of the oscillation of $p_{D}$ is $41.1 \pm 1.0\; \mu s$. This period related to the band gap and the theoretical value is $40.8\; \mu s $. From Fig.~\ref{f2}(a), we can see that the amplitude, or the contrast $C(t_{OL})$, decreases with the increase of $t_{OL}$, where
\begin{equation} \label{e2}
p_D(t_{OL})=[1+C(t_{OL})\cos (\omega t_{OL}+\phi)]/2.
\end{equation}

We defined a characteristic time $\tau$, which corresponds to the time when the $C(t_{OL})$ decreases to $1/e$. Temperature can affect the length of $\tau$.

\subsubsection{Contrast decay mechanisms }

To improve RI's coherent time and performance, we should analyze the mechanisms that cause RI signal attenuation. By solving the Gross-Pitaevskii equation(GPE), which considers the mechanism that may lead to decay, we can get the process of contrast decay in theory. In Fig.~\ref{f2}(b), The following mechanisms are introduced in turn: the effect of the imperfection of the $\pi/2$ pulse (brown dashed line), inhomogeneity of laser wavefront (blue dotted line), the transverse expansion caused by the many-body interaction (blue dashed line), laser intensity fluctuation (the dash-dotted line), and the thermal fluctuations (the orange solid line). Fig.~\ref{f2}(b) illustrates that the theoretical (the orange solid line) and experimental (black dots) curves of the final result are very consistent.

\subsubsection{An echo-Ramsey interferometer with motional Bloch states of BECs}
In order to extend the coherence time $\tau$, we proposed a quantum echo method. The echo process refers to a designed $\pi$ pulse ($\hat L(\pi)$) that flips the atomic populations of the two bands. So the evolution operator of Echo-RI is $\hat L(\pi/2)[\hat Q(t_{OL}/2n) \hat L(\pi)\hat Q(t_{OL}/2n)]^{n} \hat L(\pi/2)$, where $n$ is the number of the $\pi$ pulse inserted between the two $\pi/2$ pulses.

\begin{table}
\centering
\caption{The effects for the contrast decay.}
\begin{tabular}{ccc}
\hline
{Decay factor}      & {Beam inhomogeneity}                     & {Echo recovery} \\\hline
{$Dephasing$}                & {Momentum dispersion}                    & {Yes} \\

{$Collision$}                & {Unbalance of population}                    & {Yes} \\
{$Decoherence$}              & {fluctuation}                & {No} \\

\hline
\end{tabular}
\label{table:ram}
\end{table}

Fig.~\ref{f2}(c) illustrates the characteristic time $\tau$ for different $n$ and temperatures. And the effects for the contrast decay are listed in Table.~\ref{table:ram}. It can be seen from Fig.~\ref{f2}(c) that the interferometer with the longest characteristic time (14.5 ms) was obtained when $n\ge 6$ and $T=50$ nK.

\section{\label{sec:resolution}Enhanced resolution by multimode scheme}

As an essential indicator, the resolution evaluates the performance of interferometers. The resolution is theoretically restricted to shot-noise limit, or sub-shot noise limit~\cite{PhysRevLett.102.100401,RN256}, however, it will decay easily due to other experimental noises, with those upper limits beyond reach. Therefore, we have made several efforts to increase the resolution in practice. Improvement of the phase resolution was accomplished by a noise-resilient multi-component interferometric scheme. With the relative phase of different components remaining stable, increasing the number of paths could sharpen the peaks in the interference fringes, which leads to a resolution nearly twice compared with that of a conventional double-path two-mode interferometer with hardly any attenuation in visibility. Moreover, improvement of the momentum resolution is fulfilled with optical lattice pulses. Under the condition of 10 $\rm{E_R}$ OL depth, atomic momentum patterns with interval less than the double recoil momentum can be achieved, exhibiting 10 main peaks, respectively, where the minimum one we have given was 0.6 $\hbar k_L$. The demonstration of these techniques is shown in the next four subsections.

\subsection{Time evolution of two-component Bose-Einstein condensates with a coupling drive}

For the multicomponent interferometer, it is necessary to study the interference characteristics of multi-component ultra-cold atoms.
Here we introduced a basic method to deal with this problem, which simulates the time evolution of the relative phase in two-component Bose-Einstein condensates with a coupling drive~\cite{PhysRevA.64.015602}.

We considered a two-component Bose-Einstein condensate system with weak nonlinear interatomic interactions and coupling drive. In the formalism of the second quantization, the Hamiltonian of such a system can be written as
\begin{equation}
\hat{H}=\hat{H}_1+\hat{H}_2+\hat{H}_{int}+\hat{H}_{driv},
\end{equation}
\begin{equation}
 \hat{H}_i=\int dx \Psi_i^\dag(x)[-\frac{\hbar^2}{2m}\nabla^2+V_i(x)+U_i(x)\Psi_i^\dag(x)\Psi_i(x)]\Psi_i(x),
\end{equation}
\begin{equation}
\hat{H}_{int}=U_{12}\int dx \Psi_1^\dag(x)\Psi_2^\dag(x)\Psi_1(x)\Psi_2(x),
\end{equation}
\begin{equation}
\hat{H}_{driv}=\int dx [\Psi_1^\dag(x)\Psi_2(x)e^{i\omega_{rf}t}+\Psi_1(x)\Psi_2^\dag(x)e^{-i\omega_{rf}t}],
\end{equation}

where $i=1$ and $2$.

Then the interference between two BEC's is
\begin{equation}
I(t)=\frac{1}{2}N+\frac{1}{2}(N_1-N_2)\cos{\omega_{rf}t}+\frac{1}{2}e^{-A(t)}\sin{\omega_{rf}t}\mathcal{R}(t).
\end{equation}

Previous analysis can be used to simulate the time evolution of the relative phase in two-component Bose-Einstein condensates with a coupling drive, as well as to study the interference of multi-component ultra-cold atoms. This simulation would help to construct a multimode interferometer of a spinor BEC (see Subsecs.~\ref{sec:DoublePath}).

\subsection{Parallel multicomponent interferometer with a spinor Bose-Einstein condensate}

\begin{figure}
\begin{center}
\includegraphics[width=0.6\linewidth]{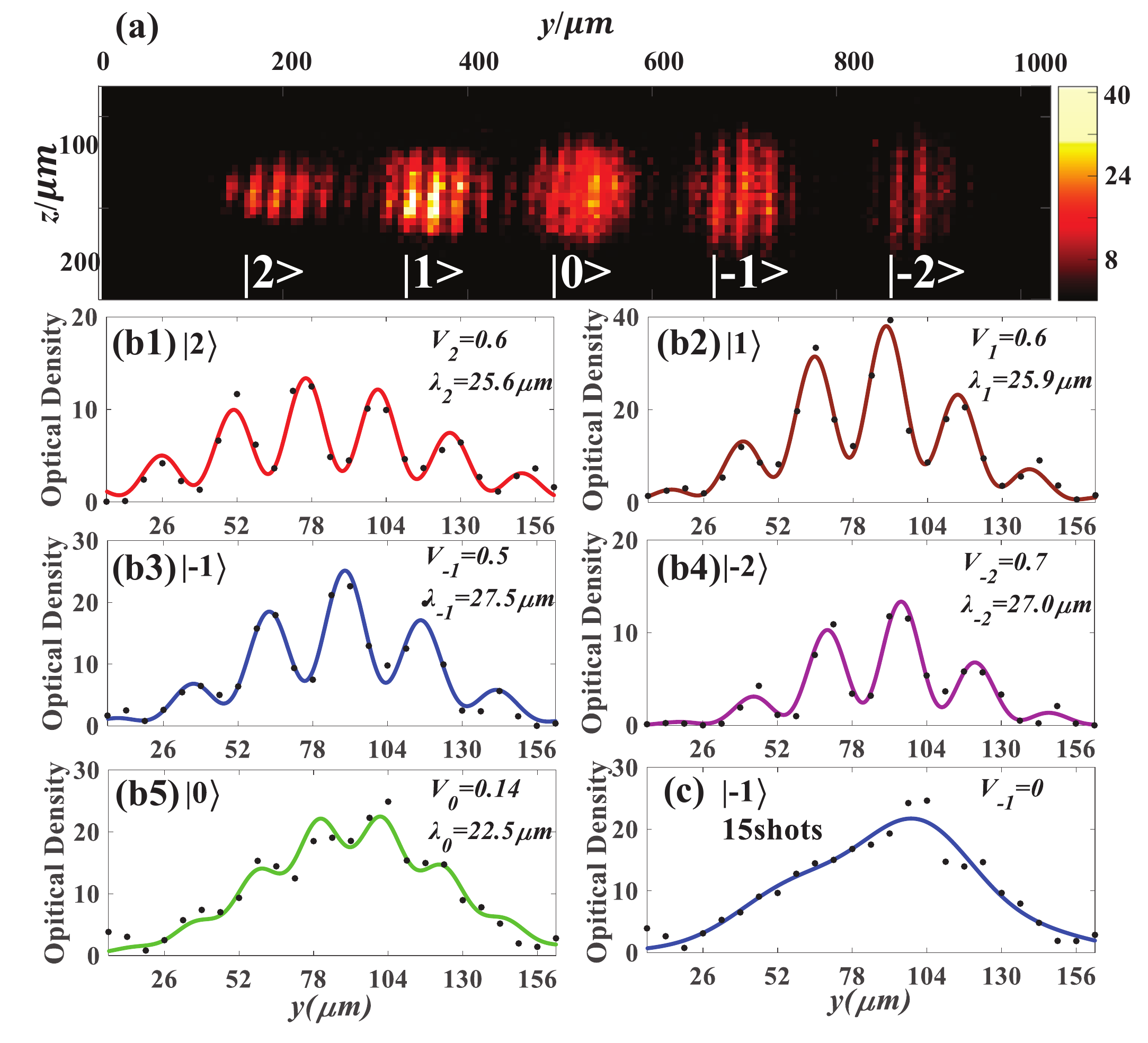}
\end{center}
\caption{
(a) One typical interference picture. These spatial interference fringes come from the five sub-magnetic states of $\left|F=2\right\rangle$ hyperfine level.
(b1-5) Density distributions corresponding to different sub-magnetic components respectively, where the points are the experimental data and the curves are fitting results according to the empirical expression~\cite{13Simsarian2000,13machluf2013coherent,12margalit2015self}.
(c) Average of 15 consecutive experimental shots with a visibility reduction to zero for the chosen state $\left|m_F=-1\right\rangle$.
Reproduced with permission from Ref.~\cite{Tang2019a}.
}
\label{fig:parallel_Fig1}
\end{figure}

\begin{figure}
\begin{center}
\includegraphics[width=0.6\linewidth]{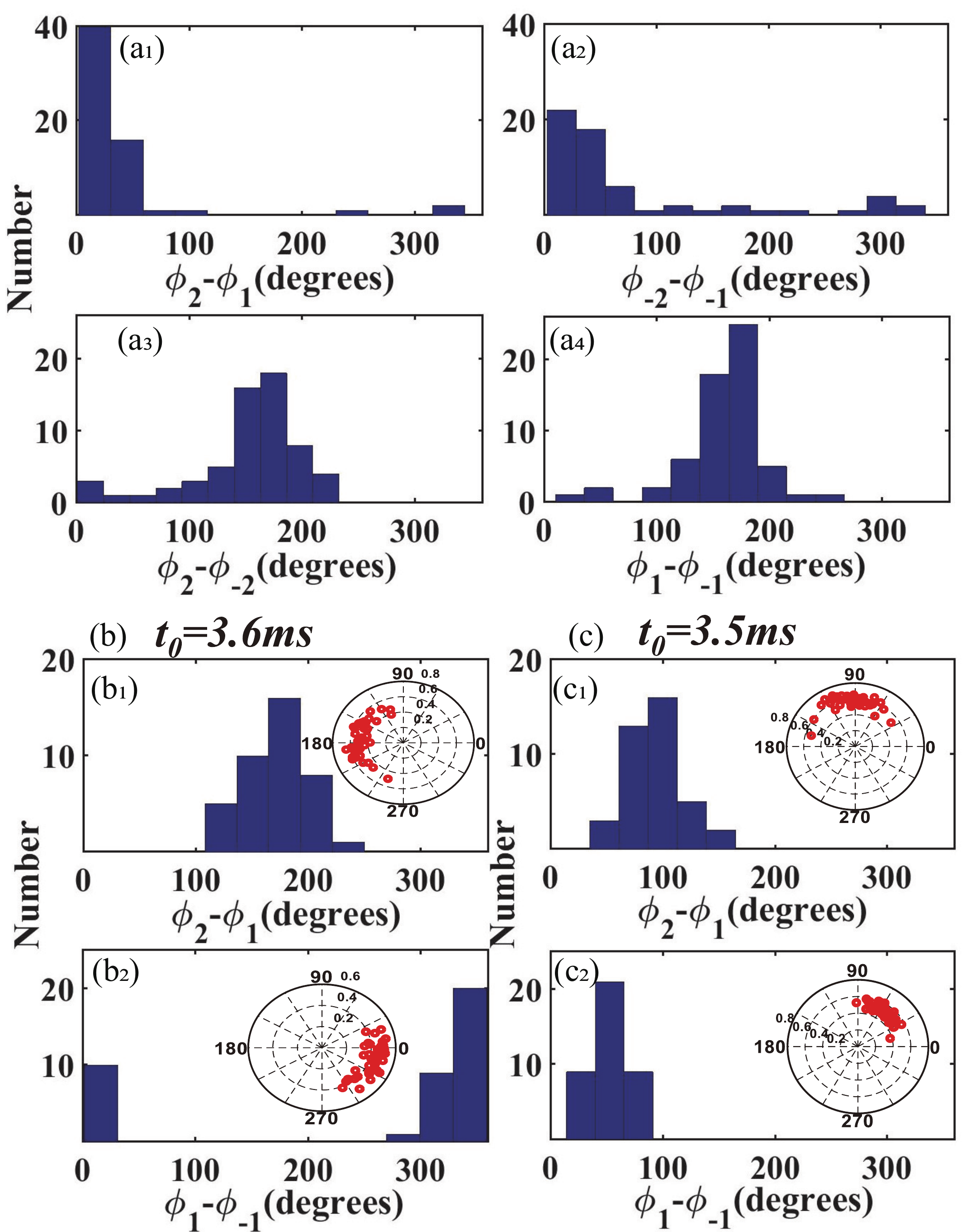}
\end{center}
\caption{
(a1)-(a4) Histograms of relative phases distributions $(\phi_2-\phi_1,\phi_{-2}-\phi_{-1},\phi_2-\phi_{-2}, and\; \phi_1-\phi_{-1})$ respectively. These relative phases show good reproducibility, for the first two are concentrated at about $0^o$, while the latter two are concentrated at about $180^o$ in 61 consecutive experimental shots;
(b) Relative phase distributions of 41 consecutive experimental shots with $t_0=3.6$ ms. Distributions of relative phases $\phi_2-\phi_1$ and $\phi_1-\phi_{-1}$ are shown in (b1) and (b2);
(c)When $t_0= 3.5$ ms, distributions of relative phases $\phi_2-\phi_1$ and $\phi_1-\phi_{-1}$ are shown in (c1) and (c2). The polar plots of relative phase vs visibility (shown as angle vs radius) are shown as these insets, respectively, where the value of visibility is an average of the visibility involved in calculation.
Reproduced with permission from Ref.~\cite{Tang2019a}.
}
\label{fig:parallel_Fig2}
\end{figure}

Revealing the wave-particle duality, Young's double-slit interference experiment plays a critical role in the foundation of modern physics. Other than quantum mechanical particles such as photons or electrons which had been proved in this stunning achievement, ultra-cold atoms with long coherent time have got the potential of precision measurements when utilizing this interferometric structure. Here we have demonstrated a parallel multi-state interferometer structure~\cite{Tang2019a} in a higher spin atom system~\cite{29PRLinterferency,30RevModPhys,31Andrews637}, which was achieved by using our spin-2 BEC of $^{87}\rm{Rb}$ atoms.

The experimental scheme is described as following. After the manufacture of Bose-Einstein condensates in an optical-magnetic dipole trap, we switched off the optical harmonic trap and populate the condensates from $\left|F=2, m_F=2\right\rangle$ state to $\left|m_F=2\right\rangle$ and $\left|m_F=1\right\rangle$ sub-magnetic level equally. After the evolution in a gradient magnetic field for time $t_1$, these two wave packets were converted again into multiple $m_F$ states ($m_F=\pm 2,\pm 1,0$) as our spin states, leading to the so-called parallel path. All these states were allowed to evolve for another period time $t_2$, then the time-of-flight (TOF) stage $t_3$ for absorption imaging. Spatial interference fringes had been observed in all the spin channels. Here, we used the technique of spin projection with Majorana transition ~\cite{21Xiuquan2006,46Majorana,23XiaLin2008} by switching off the magnetic field pulses nonadiabatically to translate the atoms into different Zeeman sublevels. The spatial separation of atom cloud in different Zeeman states was reached by Stern-Gerlach momentum splitting in the gradient magnetic field.

A typical picture after 26 ms TOF is shown in Fig.~\ref{fig:parallel_Fig1}(a). Fig.~\ref{fig:parallel_Fig1}(b1-b5) are the density distributions for each interference fringes. To reach the maximal visibility, we studied the correlation between the interference fringes' visibility and the time interval applying Stern-Gerlach process. Though separated partially, the interfering wave pockets must overlap in a sort of way. The optimal visibility was about 0.6, corresponding to $t_1 = 210\;\rm{\mu s}$ and $t_2=1300\;\rm{\mu s}$. We also measured the fringe frequencies of different components, which exhibited a weak dependence on $m_F$.

Special attention is required in Fig.~\ref{fig:parallel_Fig1}(c). After an average of 15 consecutive CCD shots in repeated experiments, the interference fringe almost disappeared for the chosen state $\left|m_F=-1\right\rangle$. This result manifested the phase difference between the two copies of each component in every experimental run is evenly distributed. The poor phase repeatability could be attributed to uncontrollable phase accumulation in Majorana transitions.

However, the relative phase across the spin components remained the same after more than 60 continuous experiments, just as Fig.~\ref{fig:parallel_Fig2}(a) illustrates. Furthermore, evidence has been spotted that the relative phase can be controlled by changing the time $t_0$ before the first Majorana transition, as shown in Fig.~\ref{fig:parallel_Fig2}(b)(c), paving a way towards noise-resilient multicomponent parallel interferometer or multi-pointer interferometric clocks~\cite{12margalit2015self}.

\subsection{\label{sec:DoublePath}Implementation of a double-path multimode interferometer using a spinor Bose-Einstein condensate}

\begin{figure}
\begin{center}
\includegraphics[width=0.6\linewidth]{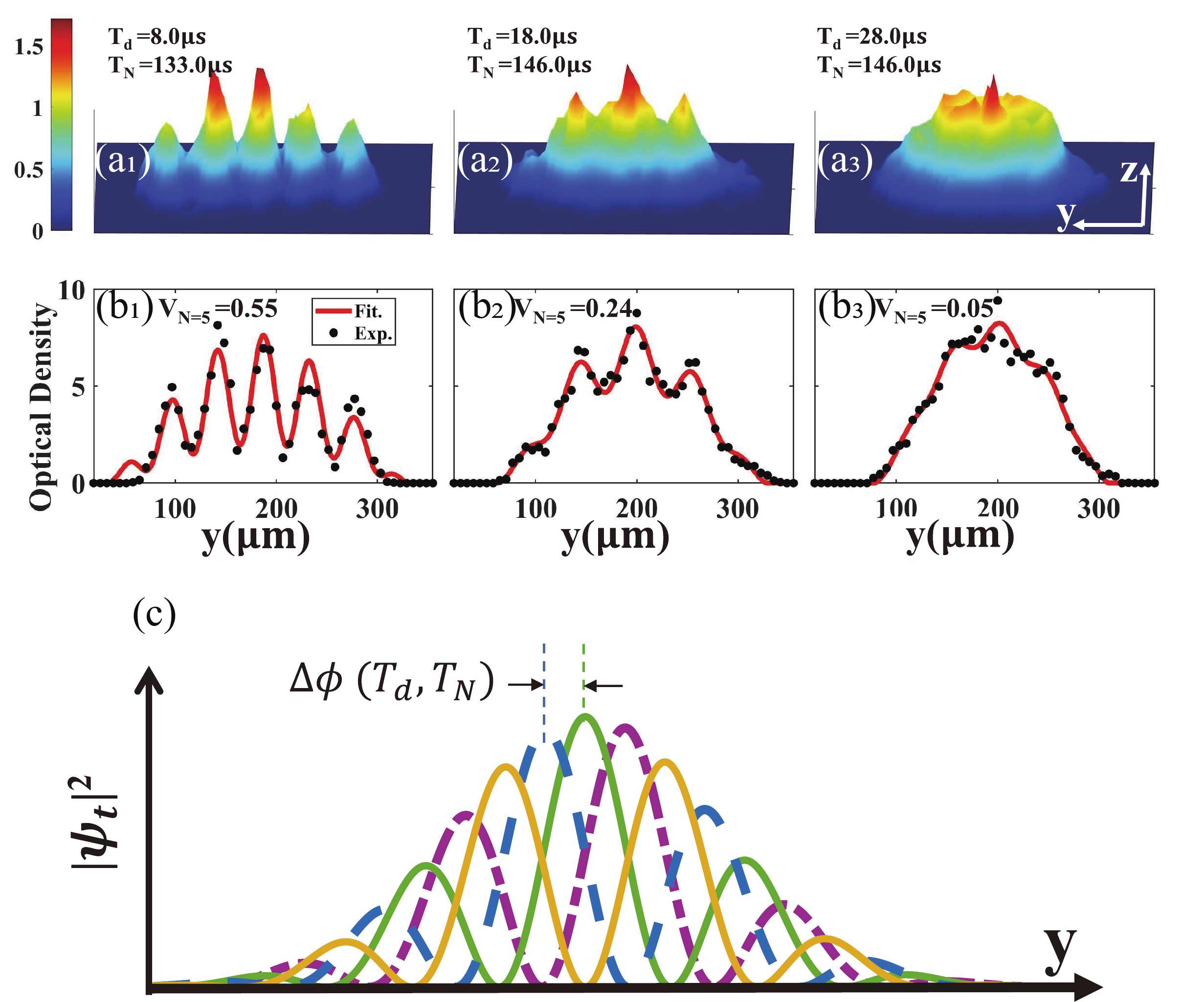}
\end{center}
\caption{
(a1-a3) Single-shot spatial interference pattern with five interference modes after TOF = 26 ms. Fringes of each mode are (a1)in phase (a2)partially in phase (a3)complementary in space.
(b1-b3) Black points are the experimental data by integrating the image in panels (a1-a3) along the z direction. Red solid lines are fitted by Thomas-Fermi Distribution~\cite{13machluf2013coherent}. Visibilities are 0.55, 0.24, and 0.05, respectively. (c) Schematic of the spatial interference image. $\Delta \phi(T_d,T_N)$ is the relative phase between adjacent mode fringes. The fringe in each color represents the interference between the two wave packets of a single mode.
Reproduced with permission from Ref.~\cite{PhysRevA.101.013612}.
}
\label{fig:double_Fig1}
\end{figure}

\begin{figure}
\begin{center}
\includegraphics[width=0.6\linewidth]{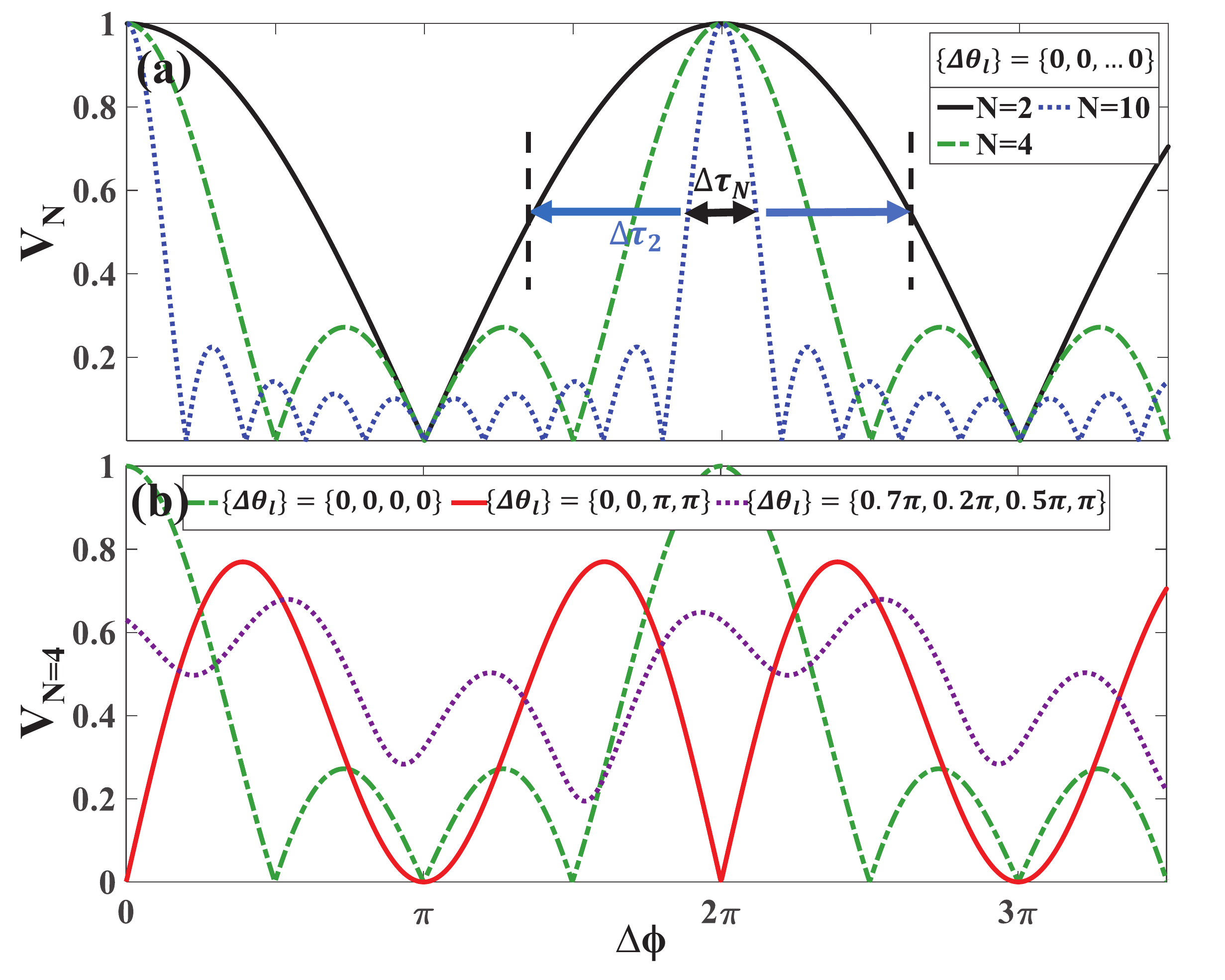}
\end{center}
\caption{
Dependence of the visibility on the number of modes N and initial relative phase $\phi_{m_F}$ of the same mode in two paths. (a) Dependence on N in a situation that $\phi_{m_F}$ are all zero. FWHM of the N-mode fringe is 2/N times that of the two-mode fringe. (b) Dependence on $\phi_{m_F}$ using $N =4$ as an example. The green dashed line, red solid line, and purple dotted line show the fringes with $(\phi_1,\phi_2,\phi_3,\phi_4)=(0,0,0,0)$, $(0,0,\pi,\pi)$, and $(0.7\pi,0.2\pi,0.5\pi,\pi)$, respectively.
Reproduced with permission from Ref.~\cite{PhysRevA.101.013612}.
}
\label{fig:double_Fig2}
\end{figure}

The experiment described above was achieved by Stern-Gerlach momentum splitting, separating the wave pockets in different spin states or Zeeman sub-magnetic states in space. The conclusion that relative phases across the spin components remain stable gives us an inspiration to carry on the double-path multimode matter wave interferometer scheme. With the number of paths increased, it will suppress the noise and improve the resolution~\cite{7.4weitz1996multiple,7.3three-path,7.2multiphoton,7.5PhysRevA.87.033607,8paul2017measuring,10petrovic2013multi} compared with the conventional double-path single-mode structure. The results show that resolution of the phase measurements is increased nearly twice in time domain interferometric fringes~\cite{PhysRevA.101.013612}.

The experimental procedure is similar to the previous one. The major difference lies in the splitting stage, during the optical harmonic trap participating in the preparation of the condensates is not going to switch off until the TOF stage, thus the Stern-Gerlach process in the gradient magnetic field mentioned above cannot significantly split the wave packets. With different momentum atomic clouds are spatially separated only for tens of nanometers, approximately 1\% of the BEC size, thus well overlapped~\cite{13machluf2013coherent}. As a result, multi-modes from two paths will interfere in one region instead of five. Another difference lies in the second spin projection with non-adiabatic Majorana transition. Here we replace it with a radio frequency pulse for its higher efficiency as a 1 to 5 beam splitter, although that we still use it to transfer the initial condensates into $\left|m_F=2\right\rangle$ and $\left|m_F=1\right\rangle$ sub-magnetic levels. The performance of Majorana transition is better than RF pulse as a 1 to 2 beam splitter. There are also some changes with experimental parameters that count a little and we would not discuss them here.

Hence the global view of our interferometer is as follows: The magnetic sublevels are considered as modes in the interferometer, each has its own different phase evolution rates in gradient magnetic field. The double path configuration is made up of Majorana transition as well as the evolution of the first two $m_F$ superposed states during time $T_d$, makes up (path I, path II). RF pulse leading to the multiple $m_F$ superposed states together with their evolution in time $T_N$ forms the multi-modes configuration. During the TOF stage, atomic clouds expand and interfere with each other. Owing to the different state-dependent phase evolution rate $\omega_{m_F}^{(I,II)}$, the absorption image shows something more than spatial interference fringes, which is a periodic dependence of the visibility on phase evolution time as the function $V_N (T_d,T_N)$. We refer to it as the time domain interference.

Fig.~\ref{fig:double_Fig1}(a) shows a group of absorption images with various combinations of $T_d$, $T_N$. The observed fringe is a superposition of the interference fringes of different modes. Consequently, the visibility depends on the relative phase $\Delta \phi(T_d,T_N)$ between the interference fringes of each mode [Fig.~\ref{fig:double_Fig1}(c)] and can also be modulated.

By carefully analyzing with expression $V_N (T_d,T_N)=\langle \Psi^{(I)}|\Psi^{(II)}\rangle$[34], we can acquire the expression of the relative phase between two adjacent components:
\begin{align}\label{eqn:phi}
	\begin{split}
		\Delta\phi(T_d,T_N)
		&=(\Delta\phi_{m_F}-\Delta\theta_{m_F-1})\\
		&=\Delta \omega T_N+\Delta\theta
	\end{split}
\end{align}
where $\Delta \omega$ is the relative phase evolution rate between the two paths, $\Delta \theta$ is the relative initial phase introduced through the double path stage $T_d$. Yet we have already demonstrated that the visibility $V_N$ is modulated with the period $2\pi/\Delta \omega$ along with how the time domain fringe emerges theoretically.

A remarkable feature of the multi-modes interferometer is the enhancement of resolution, which is defined as (fringe period)/(full width at half maximum). We have investigated the resolution of the time domain fringe experimentally and theoretically. It can be influenced by parameters like modes number and initial phase, which is $R(N,\phi_{m_F})$. $\phi_{m_F}$ refers to the initial relative phase of $m_F$ states accumulated in double paths $T_d$.

Fig.~\ref{fig:double_Fig2} is the numerical results considering an arbitrary number of modes. Fig.~\ref{fig:double_Fig2}(a) is under the condition that the phases $\phi_{m_F}$ are all the same for any modes. In that case, if we denote $\Delta \omega T_N=2n\pi/N$, then the visibility achieves $V_N=1$ when n is the multiple of N and a major peak is observed in this case. A remarkable feature of our interferometer is the enhancement of resolution by $N/2$ times without any changes in visibility nor periodical time. It is the harmonics that cause the peak width to decrease with the number of modes increasing in this case~\cite{3.4Pikovski2017clock}. Fig.~\ref{fig:double_Fig2}(b) indicates $\phi_{m_F}$ varies from mode to mode for comparison. Neither the maximum visibility $V_N=1$ nor the minimum could be reached. Meanwhile, the time domain fringe shows more than one main peak in one period. Therefore, the initial phase $\phi_{m_F}$ needs to be well controlled to achieve the highest possible visibility and clear interference fringe in the time domain.

We also experimentally study the time domain fringes. The experimental data (not depicted here) coincides with the numerical results of Fig.~\ref{fig:double_Fig2}(b) red line, testifying its superiority to the resolution of the phase measurement. Moreover, the relative phase evolution rate $\Delta \omega$ can be controlled by adjusting the difference between the two paths accumulated in $T_d$ stage~\cite{1dalfovo1999theory,13Simsarian2000,7fort2001spatial}. With enhanced resolution, the sensitivity of interferometric measurements of physical observables can also be improved by properly assigning measurable quantities to the relative phase between two paths, as long as the modes do not interact with each other~\cite{14PhysRevA.67.053803,11PhysRevA2017multimode,12margalit2015self}.

\subsection{Atomic momentum patterns with narrower interval}

\begin{figure}
\begin{center}
\includegraphics[width=0.6\linewidth]{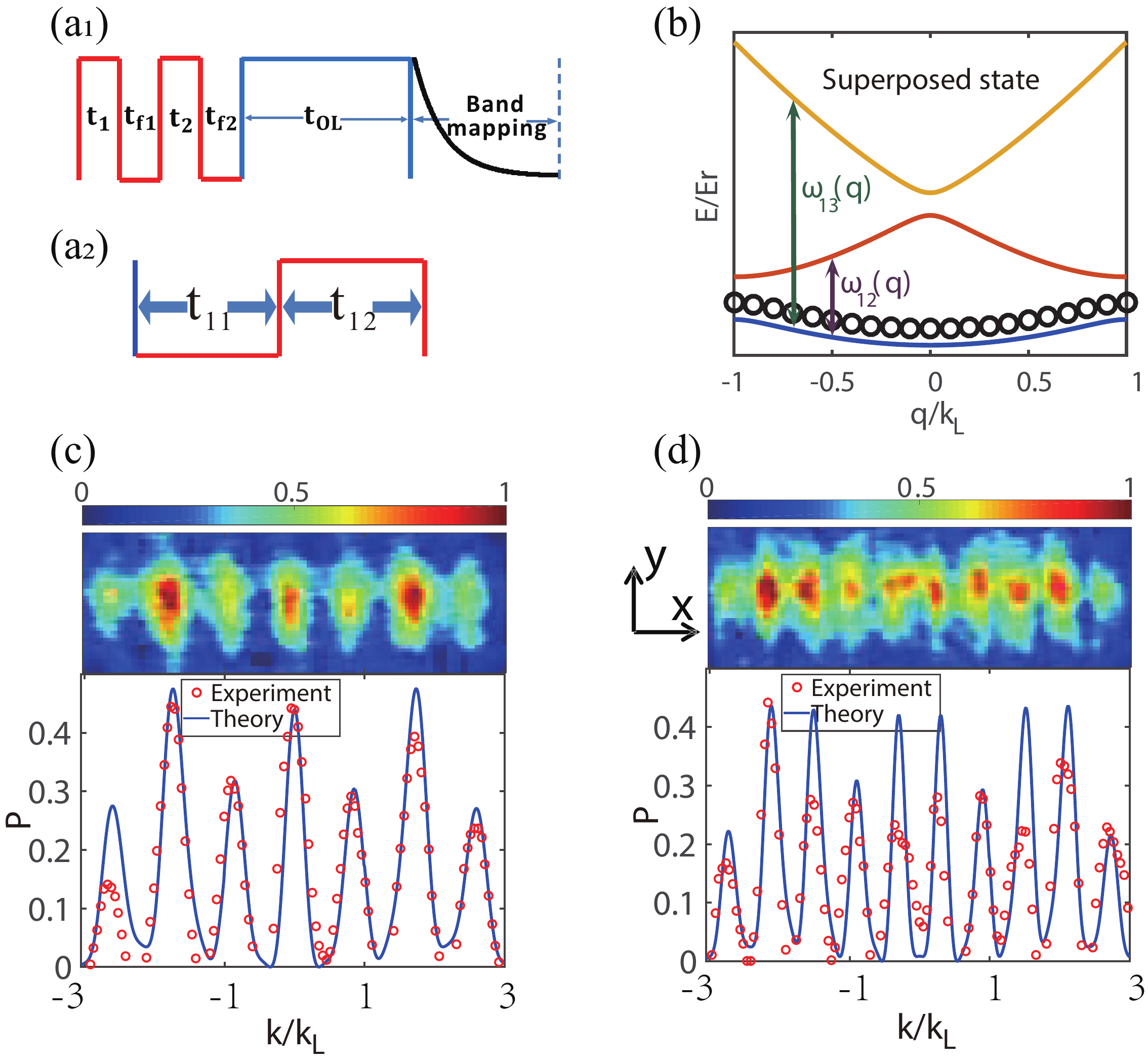}
\end{center}
\caption{
(a) Shortcut method for loading atoms: (a1) after the first two pulses and the $30$ ms holding time in the OL and the harmonic trap, the state becomes the superposition of the Bloch states in S-band with quasi-momenta taking the values throughout the FBZ, and is denoted by $\left| {\psi \left( 0 \right)} \right\rangle$. Then $1$ms band mapping is added.
(a2) The single pulse acted on the superposed state $\left| {\psi \left( 0 \right)} \right\rangle$.
(b) the superposed Bloch states of S-band spreading in the FBZ (black circles).
The top Patterns in (c) and (d) are the TOF images in experiments.
The lower part of (c) and (d) depicts the atomic distributions in experiments (red circles) and theoretical simulations (blue solid lines). There are seven peaks in (c) and ten peaks in (d) with $q=\pm 3\; \hbar k_L$.
Reproduced with permission from Ref.~\cite{RN112}.
}
\label{fig:patterns}
\end{figure}

For ultra-cold atoms used in precise measurement, improving the precision of momentum manipulation is also conducive to improving the measurement resolution.
The method to get atomic momentum patterns with narrower interval has been proposed and verified by experiments~\cite{RN112}. Here we applied the shortcut pulse to realize the atomic momentum distribution with high resolutions for a superposed Bloch states spreading in the ground band of an OL.

While difficult to prepare this superposition of Bloch states, it can be overcome by the shortcut method. First, the atoms are loaded in the superposition of S- and D-bands $(|S,q=0\rangle+|D,q=0\rangle)/\sqrt{2}$, where $q$ is the quasi-momentum. Fig.~\ref{fig:patterns}(a1) depicts the loading sequence. The atoms in S and D bands Collision between atoms in S and D bands will cause the atoms to gradually transfer to S band with non-zero quasi-momentum. After $30$ ms, as shown in Fig.~\ref{fig:patterns}(b), atoms cover the entire ground band from $q=-\hbar k_L$ to $\hbar k_L$.

The momentum distribution of the initial state is a Gaussian-like shape. After an OL standing-wave pulse, which is similar to that in the shortcut process, the different patterns with the narrower interval can be obtained. The standing-wave pulse sequence is shown in Fig.~\ref{fig:patterns}(a2).
Fig.~\ref{fig:patterns}(c) and (d) show the different designs for patterns of multi modes with various numbers of peaks under OL depth 10 $E_r$, where the top figures are the absorption images after the pulse and a $25$ ms TOF.
The red circles are the experimental results of the atomic distribution along the x-axis from the TOF images. These results are very close to the numerical simulation result (blue lines). For the numerical simulation, we can get the initial superposition of states by fitting the experimental distributions (Fig.~\ref{fig:patterns}(b)). Fig.~\ref{fig:patterns}(c) and (d) depict the atomic momentum distribution with resolutions of $0.87\; \hbar k_L$ interval (seven peaks within $q=\pm 3\;\hbar k_L$) and $0.6\; \hbar k_L$ (ten peaks within $q=\pm 3\;\hbar k_L$) interval, respectively.

The superposed states with different quasi-momenta in the ground band cause the narrow interval (far less than double recoil momentum) between peaks, which is useful to improve the resolution of atom interferometer~\cite{30RevModPhys}.

\section{\label{sec:noises}Enhanced resolution by removing the systematic noise}
Noise identification as well as removal is crucial when extracting useful information in ultra-cold atoms absorption imaging. In general concept, systematic noises of cold atom experiments originate from two sources, one is the process of detection, such as optical absorption imaging; the other is the procedure of experiments, such as the instability of experimental parameters. Here we provided an OFRA scheme, reducing the noise to a level near the theoretical limit as $1/\sqrt{2}$ of the photo-shot noise. When applying the PCA, we found that the noise origins, which mainly come from the fluctuations of atom number and spatial positions, much fewer than the data dimensions of TOF absorption images. These images belong to BECs in one dimensional optical lattice, where the data dimension is actually the number of image pixels. If the raw TOF data can be preprocessed with normalization and adaptive region extraction methods, these noises can be remarkably attenuated or even wiped out. PCA of the preprocessed data exhibits a more subtle noises structure. When we compare the practical results with the numerical simulations, the few dominant noise components reveal a strong correlation with the experimental parameters. These encouraging results prove that the OFRA as well as PCA can be a promising tool for analysis in interferometry with higher precision~\cite{Segal, Chiow-11}.

\subsection{Optimized fringe removal algorithm for absorption images}

Optical absorption imaging is an important detection technique to obtain information from matter waves experiments. By comparing the recorded detection light field with the light field in the presence of absorption, we can easily attain the atoms' spatial distribution. However, due to the inevitable differences between two recorded light field distributions, detection noises are unavoidable.

\begin{figure}[htbp]
\centering
\includegraphics[width=0.5\linewidth]{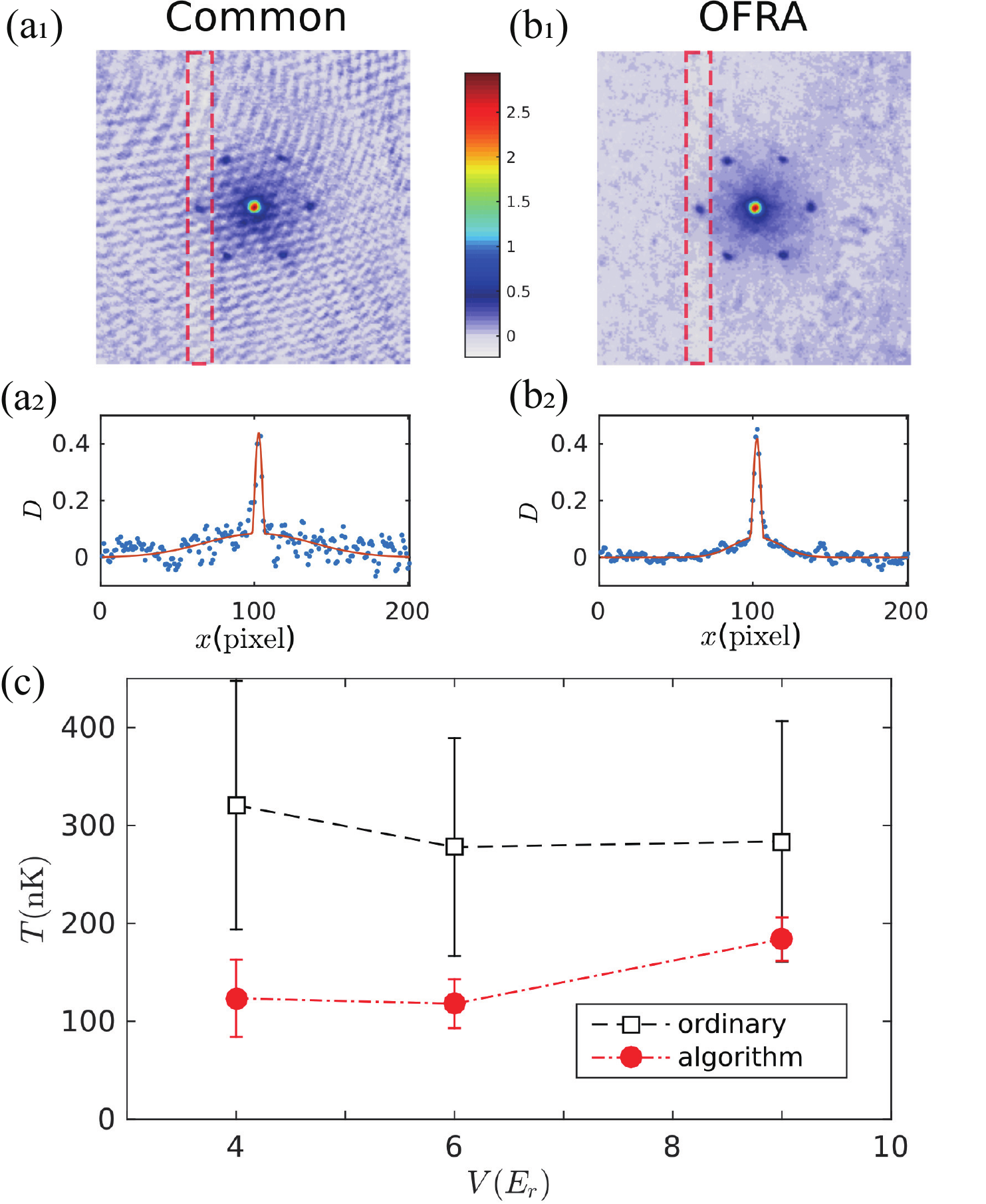}
\caption{Comparison between ordinary method and OFRA method.
The integral of the atomic distribution in the red box in (a1) and (b1) correspond to (a2) and (b2). The atomic distribution (blue dots) is fitted by a bi-modal function to extract the temperature of atoms, which is shown in (c).
Reproduced with permission from Ref.~\cite{Nlx2018}.}
\label{fringe}
\end{figure}

Therefore, we have demonstrated an OFRA scheme to generate an ideal reference light field. With the algorithm, noise generated by the light field difference could be eliminated, leading to a noise close to the theoretical limit~\cite{Nlx2018}. The OFRA scheme is based on the PCA, we confirmed its validity by experiments of triangular optical lattices. The experimental configuration has been described in our prior work~\cite{Zhou2018,Jin2019}. When the experiment was in process, the depth of the lattice was adiabatically raised to a final value, followed by a hold time of 20 ms to keep the atoms in the lattice potential before the optical absorption imaging. There are several parameters to characterize the triangle lattice system~\cite{Becker2010}, among which the visibility, the condensate fraction, and the temperature matter. Fig.~\ref{fringe} (a2) and Fig.~\ref{fringe} (b2) are bimodal fitting to the scattering peaks by summing up the atomic distribution within the red box in the direction perpendicular to the center. Here Fig.~\ref{fringe} (a) stands for the common way of calculation and 9(b) for the OFRA. The bi-modal curve is composed of two parts: a Gaussian distribution for the thermal component and an inverse parabola curve for the condensed atoms. For each part, the column densities along the imaging axis can be written as

\begin{eqnarray}
n_{th}(x) &=& \frac{n_{th}(0)}{g_2(1)}g_2[\exp(-(x-x_0)^2/\sigma_T^2)],\\ \nonumber
n_{c}(x) &=& n_{c}(0)\max[1-\frac{(x-x_0)^2}{\chi^2}].
\end{eqnarray}

In the formula there are 5 parameters accounting for the bi-mode fitting, the amplitude of two
components $n_{th}(0)$ and $n_{c}(0)$, the width of two components $\sigma_T$, $\chi$ and the center position $x_0$ of the atomic cloud. The Bose function is defined as $g_j(z) = \sum_iz^i/i^j$. In practice, we performed the least-squares fitting of $n_{th}(x)+n_{c}(x)$ to the real distribution obtained from the imaging. From the fit, we can get the atom number and width of the two components separately. Note that the measurement in Fig.~\ref{fringe} (9c) is performed at different lattice depths. For each lattice depth, 30 experiments have been performed to acquire the statistical results. The temperature is given as $T=1/2M\sigma_T^2/t_{TOF}^2/k_B$, where $M$ described the atom mass and $t_{TOF}$ width of the thermal part~\cite{PhysRevA.55.R3987}.

For the number of condensed atoms, the fitting outcome is less affected by the fringe shown in Fig.~\ref{fringe} (a). Whereas the influence of the fringe on the fitting of temperature is much more evident. The temperature is proportional to the width of the Gaussian distribution $\sigma_T$ as mentioned above. Fig.~\ref{fringe}(c) shows the temperature extracted from the TOF absorption images with and without the OFRA separately, namely Fig.~\ref{fringe}(a1) and \ref{fringe}(b1). Fig.~\ref{fringe}(a2) and \ref{fringe}(b2) are the corresponding integrated one-dimensional atomic distributions for each method. Fig.~\ref{fringe} depicts that the temperature we get with the common way of calculation has a large error of 400 nK, extraordinary higher than the initial BEC temperature of 90 nK. The turning on the procedure of lattice potential would indeed lead to a limited heating effect, nevertheless the proportion of condensed atom should be reduced significantly considering our system has been heated up by 4 times. This is still not consistent with the observation. However, the temperature is measured with much small variance at a much reasonable value if we dive into the OFRA scheme. For example, the measured temperature is 123.5 nK for a lattice depth of $V=4\;E_r$, with 183.9 nK for $V=9\;E_r$. Comparison between these two results illustrates that only by using the fringe removal algorithm we can get a reliable result, especially in the case of small atom numbers when fitting physical quantities such as the temperature.

In conclusion, with this algorithm, we can measure parameters with higher contradiction to the conventional methods. The OFRA scheme is easy to implement in absorption imaging-based matter-wave experiments as well. There is no need to do any changes to the experimental system, only some algorithmic modifications matter.

\subsection{Extraction and identification of noise patterns for ultracold atoms in an optical lattice}
Furthermore, on the basis of the absorption images after preprocessing by OFRA, the PCA method is used to identify the external noise fluctuation of the system caused by the imperfection of the experimental system. The noise can be reduced or even eliminated by the corresponding data processing program.
It makes the task more difficult that these external systematic noises are often coupled, covered by nonlinear effects and a large number of pixels. PCA provides a good method to solve this problem~\cite{LEVINE, PCANJP1, PCANJP2, PCAPRA, Segal}.

PCA can decompose the fluctuations in the experimental data into eigenmodes and provide an opportunity to separate the noises from different sources. For BEC in a one-dimensional OL, it was proved that PCA could be applied to the TOF images, where it successfully separated and recognizing noises from different main contribution sources, and reduced or even eliminated noises by data processing programs~\cite{Caoshuyang}.

The purpose of PCA is to use the smallest set of orthogonal vectors, called principal components  (PCs) to approximate the variations of data while preserving the information of datasets as much as possible. The PCs correspond to the fluctuations of the experimental system, which can help to distinguish the main features of fluctuations. In the experimental system of BECs, the data are usually TOF images. A specific TOF image $A_i$ can be represented by the sum of the average value of the images and its fluctuation:
\begin{eqnarray}
A_{i} = \bar{A} + \sum\varepsilon_{ij}P_{j},
\end{eqnarray}
Here $P_j$ is the different eigenmodes of fluctuation, $\varepsilon_{ij}$ is the weight of the eigenmodes $P_j$.

Taking the BEC experiment~\cite{Caoshuyang,Hu2018, Zhai, Zhou2018} in an OL as an example, we demonstrated the protocol of PCA for extraction noise.
A TOF image for ultra-cold atoms in experiments can be represented as a $h\times w$ matrix. The PCA progress, shown in Fig.~\ref{fig:schema}, will be applied to the images:

\begin{figure}
\centering
\includegraphics[width=12cm]{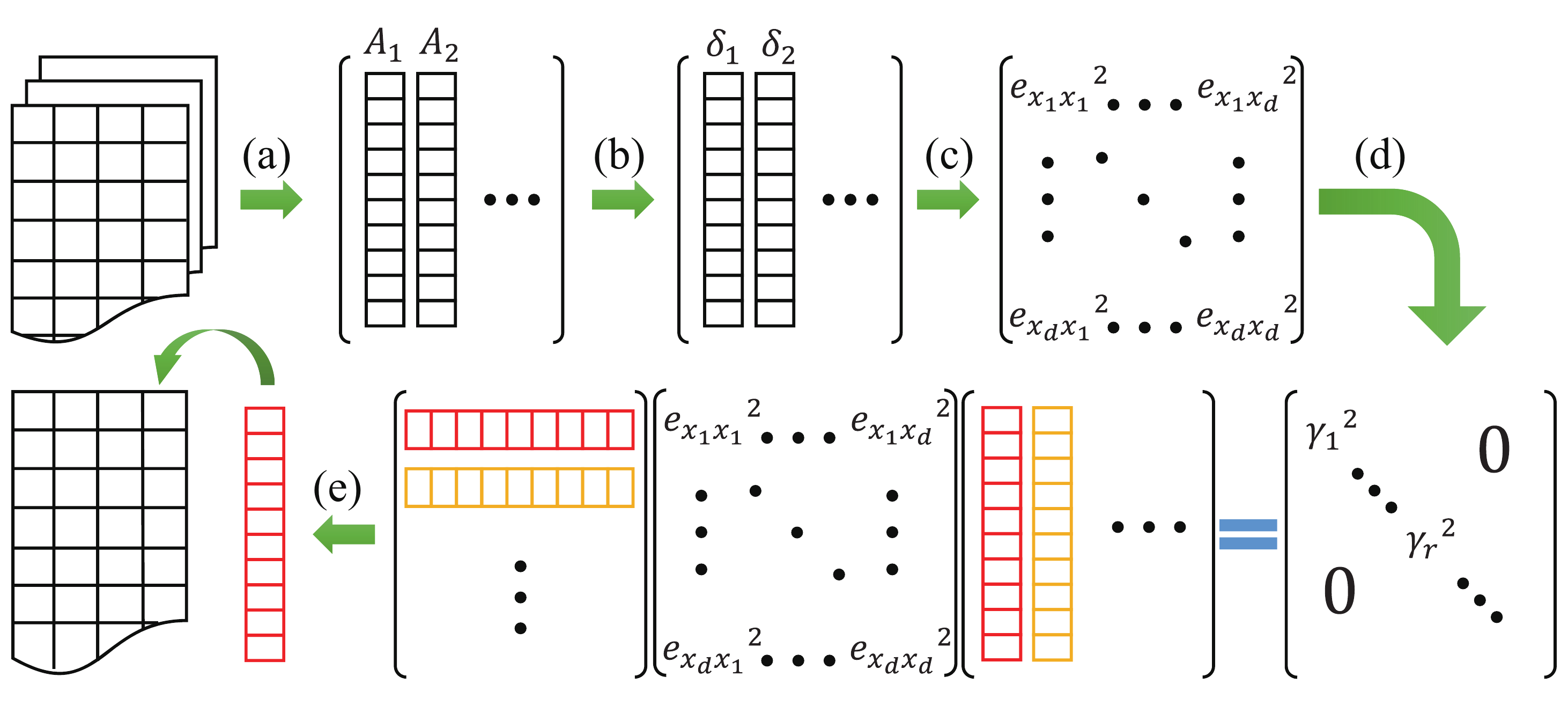}
\caption{\label{schema}Process diagram of PCA method. (a) Transform the $h\times w$ matrix (raw images) into a $1\times hw$ vector, denoted by $A_i$. And stack these vectors together. (b) Calculate the mean vector $\bar{A}= \frac{1}{n}\sum_{i=1}^{n} A_i$ and the fluctuations $\delta_i=A_i-\bar{A}$. Leaving only the fluctuation term in the matrix. (c) Stack $\delta_{i}$ together to form a matrix $X = \left[\delta_{1},\delta_{2},\cdots,\delta_{n}\right]$. Then the covariance matrix $S$ is obtained by $S = \frac{1}{n-1}X\cdot X^{T}$. (d) Decompose covariance matrix so that ${{V}^{-1}{S}{V} = {D}}$ where D is a diagonal matrix. (e) Transform eigenvectors of interest back to a new TOF image.
Reproduced with permission from Ref.~\cite{Caoshuyang}.}
\label{fig:schema}
\end{figure}

\textbf{(1)} Transform the $h\times w$ matrix into a $1\times hw$ vector, denoted by $A_i$.

\textbf{(2)} Express $A_i$ as the sum of the average value $\bar{A}$ and the fluctuation $\delta_{i}$, where $\bar{A}= \frac{1}{n}\sum_{i=1}^{n} A_i$ and $\delta_{i} = A_{i} - \bar{A}$.

\textbf{(3)} Stack $\delta_{i}$ together to form a matrix $X = \left[\delta_{1},\delta_{2},\cdots,\delta_{n}\right]$. Then the covariance matrix $S$ is obtained by $S = \frac{1}{n-1}X\cdot X^{T}$.

\textbf{(4)} Decompose covariance matrix with ${{V}^{-1}{S}{V} = {D}}$, where $V$ is the matrix of eigenvectors.

\begin{figure}
\centering\includegraphics[width=12cm]{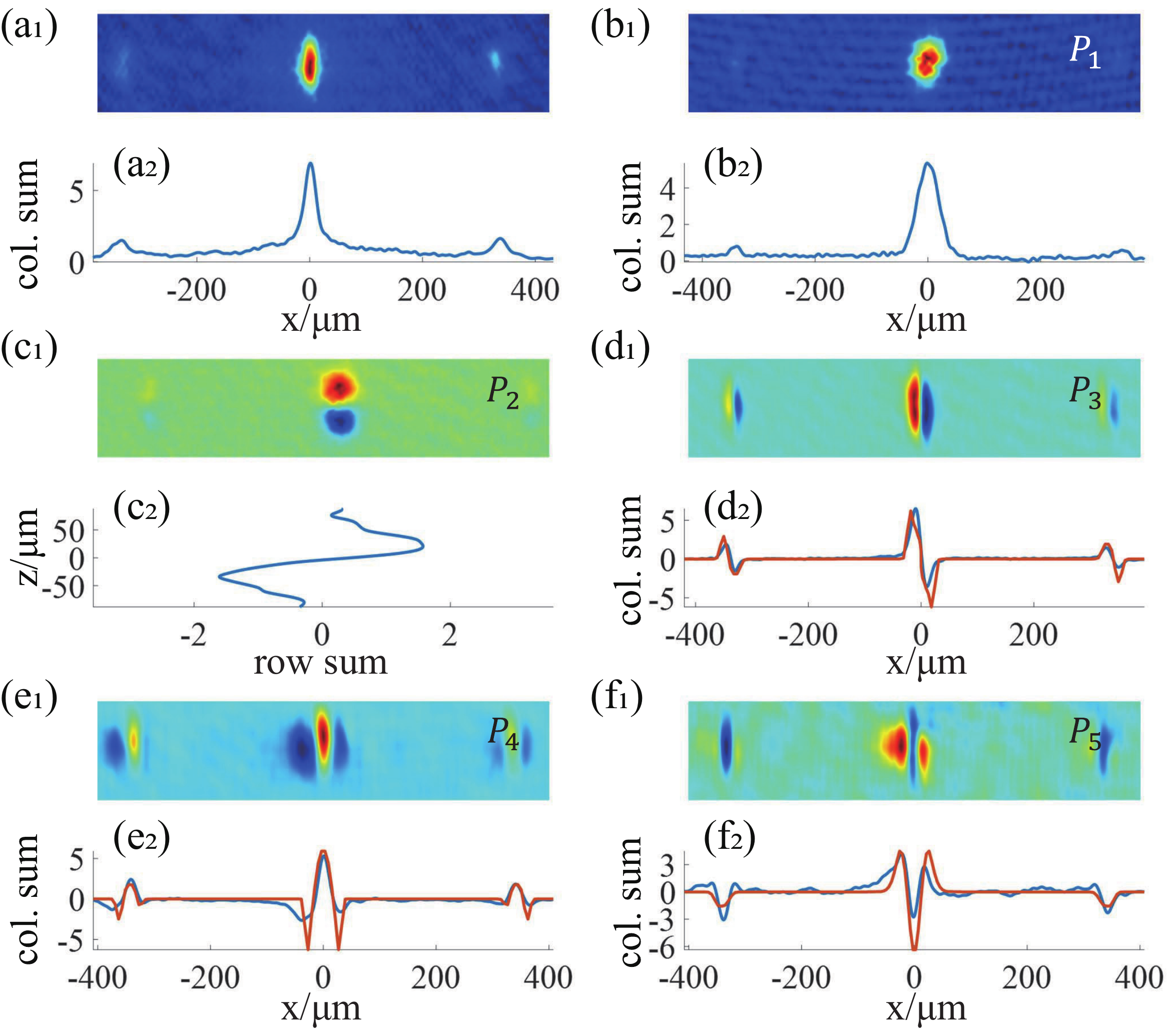}
\caption{PCA results of the TOF images.
(a) Example of a raw TOF image.
(b)-(f) correspond to the fluctuation in atom number (b), atom position (c)(d), peak width (e) and normal phase fraction (f), respectively. (a$_2$), (b$_2$), (d$_2$), (e$_2$), and (f$_2$) are the integrated results of atom distributions along x direction. (c$_2$) is for the atom distribution along z direction. The blue lines are the experimental results, and the orange lines are the simulation results.
Reproduced with permission from Ref.~\cite{Caoshuyang}.}
\label{fig:ol_pca_img}
\end{figure}

Fig.~\ref{fig:ol_pca_img} shows the PCA results of the TOF images. Fig.~\ref{fig:ol_pca_img}(b)-(f) correspond to the first to fifth PCs, respectively.

The first PC is from the fluctuation in the atom number. The normalization process can be applied to reducing the fluctuation in the atomic number. Because for a macroscopic wave function $\Psi{\left(\textbf{r}\right)}=\sqrt{N}\phi{\left(\textbf{r}\right)}$~\cite{macroscopic}, we usually concentrate on the relative density distribution, instead of the $\sqrt{N}$. After the normalization, the impact of this PC becomes very small.

The second and third PCs correspond to the position fluctuations along the $z$- and $x$-directions, respectively. The fluctuation in spatial position of the TOF images originates from the vibration of the system structure, such as the OL potential, trapping potential, and imaging system. We used a dynamic extraction method to eliminate the fluctuation in spatial position. We chose a region whose center is also the center of the density distribution. We first set a criterion to determine the center of the density distribution in the extraction area, and then used this center as the center of the new area to extract the new one. We repeated this process until the region to be extracted becomes stable.

The fourth PC is from the fluctuation in the width of the Bragg peaks in the TOF images. The final PC shown in Fig.~\ref{fig:ol_pca_img}(f1) comes from the normal phase fraction fluctuation.

By studying the first five feature images, we have identified the physical origins of several PCs leading to the main contributions. We numerically simulated this understanding using the GPE with external fluctuation terms, and got very consistent results~\cite{Caoshuyang}.
It is helpful to understand the physical origins of PCs in designing a pretreatment to reduce or even eliminate fluctuations in atom number, spatial position and other sources. Even in the absence of any knowledge of the system, the PCA method is very effective to analyze the noise, so that it can be applied to interferometers with higher precision~\cite{Segal, Chiow-11}.

\section{\label{sec:proposal}Proposal on gravity measurements}

\begin{figure*}
\centering\includegraphics[width=15cm]{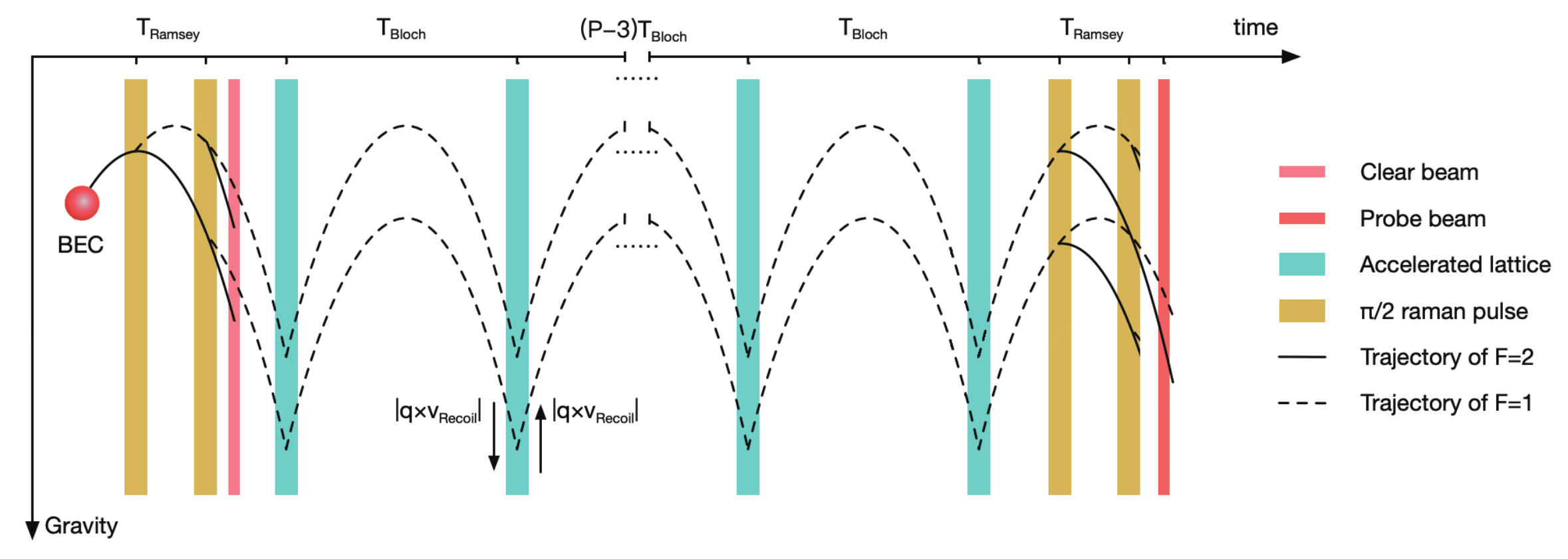}
\caption{Schematic of the experimental protocol. $^{87}\rm{Rb}$ atoms are evaporation cooled as a Bose-Einstein Condensate in the $\left|F=2, m_F=0\right\rangle$ initial state. However, they are transferred to $\left|F=1, m_F=0\right\rangle$ as the two arms of interferometry, followed by $P$ sequence of accelerate optical lattice pulse to maintain the atoms against gravity: When atoms fall to a velocity of $q\times v_{Recoil}$, they acquire a velocity of $2q\times v_{Recoil}$ upwards. The delay $T_{Bloch}$ is chosen as $T_{Bloch}= 2 q v_{Recoil}/g$ to eliminate the fall caused by gravity. Still, the probe beam should consist a laser light resonant with the $F=1$ ground state to upper levels, thus we can take absorption photos of $F=1$ population for analysis.
}
\label{fig:Gravity}
\end{figure*}

Inertia measurements~\cite{PhysRevA.80.063604,Gustavson2000,LeRN260,SchmidtRN261,PhysRevLett.107.133001,TackmannRN263,AltinRN262}, especially those for gravity acceleration $g$, have always drawn lots of attention. Until now, the performance of atom interferometry has reached a sensitivity of $8\times10^{-9}$ at 1 second~\cite{PhysRevLett.100.031101,Louchet2011}, pushing forward the determination of the Newtonian gravitational constant G~\cite{Fixler2007,PhysRevLett.100.050801} or the verification of equivalence principle~\cite{PhysRevLett.111.151102,PhysRevLett.106.151102}. Yet the bulky size of these quantum sensors strictly restricts their application for on-site measurements. Therefore, based on our previous study of ultra-cold atoms in precision measurements, we intend to precisely measure the local gravity acceleration with our $^{87}\rm{Rb}$ Bose-Einstein Condensates in a small displacement. Note that this conception bases on previous research of Perrier Clad\'{e}~\cite{PhysRevA.88.031605}.

Fig.~\ref{fig:Gravity} illustrates the protocol of this BEC gravimeter. Instead of the Mach-Zender method $(\pi/2-\pi-\pi/2)$ widely used in free-falling or atomic fountain gravimeters, we utilize the Ramsey-Bord\'{e} approach by two pairs of $\pi/2$ Raman pulses to get  a small volume. Application of the Doppler-sensitive Raman beam rather than the 6.8 GHz microwave field provides a far more efficient way to realize larger momentum splitting, which will significantly boost the interference resolution. Raman light pulse can also attain the effects of velocity distribution. Consequently, the first pair of $\pi/2$ pulses selects the initial velocity while the second pair can measure the final distribution. It should be noticed that right after the velocity selection step, a cleaning light pulse resonant with the $D_2$ line will shine on the condensate to clear away atoms remaining in $F=2$ state, leading to the two arms in interferometry.

The critical feature lies in the evolving stage between the two pairs of $\pi/2$ pulses. By periodically inverting the velocity, the two arms shall replicate parabolic trajectory in a confined volume, without any decreasing of interrogation time. Choosing appropriate parameters, displacement of ultra-cold atoms can be limited in a few centimeters, at least 1 order less than that of a Mach-Zender gravimeter. This assumption is accomplished by a succession of Bloch oscillation (BO) in a pulsed accelerated optical lattice which transfers many photon recoils to the condensates~\cite{PhysRevA.55.2989,PhysRevLett.76.4508,PhysRevLett.76.4512,PhysRevLett.92.253001}. Here the application of ultra-cold atoms instead of optical molasses~\cite{30RevModPhys} makes it more efficient when loading the atoms in the first Brillouin zone adiabatically, owing to their wave function consistency and narrow velocity distribution. This pulsed accelerated optical lattice should be manufactured along the direction of gravity with higher lattice depth, to minimize the effect of Landau-Zener Tunneling loss.

To deduce the value of g, we may scan the evolving time between the two pairs of $\pi/2$ pulses, keeping the Raman frequency of each pair of $\pi/2$ pulses fixed. When the time interval equals $2P q v_{Recoil}/g$, where $P$ is the number of pulses and $q$ is the number of recoil velocities($v_{Recoil}$) obtained by a single pulse (shown in Fig.~\ref{fig:Gravity}), the two arms are in phase and the value of g can be extrapolated. Here the absorption image is used to extract the interference information, due to the number of atoms being one order less than that obtained by the conventional method.

We also give a qualitative analysis of this compact BEC gravimeter. Besides its enormous potential in transportable instruments, prospective sensitivity maybe even better. This encouraging outlook can be attributed to a longer coherent time of ultra-cold atoms where the phase shift scales quadratically. The smaller range of movement possesses other superiorities. Systematic errors stemming from the gradients of residual magnetic fields and light fields become negligible, especially for Gouy phase and wave-front aberrations~\cite{Louchet2011,Peters2001,PhysRevA.74.052109}. Furthermore, the value of the gravity acceleration is averaged over a smaller height compared with the Mach-Zender ones. Finally, this vertical Bloch oscillation technique offers a remarkable ability to coherently and efficiently transfer photon momenta ~\cite{PhysRevLett.92.253001}, though decoherence induced by the inhomogeneity of the optical lattice must be taken into consideration.

In conclusion, we believe that this compact BEC gravimeter will have a sensitivity of a few tens of ${\mu}$Gal at least. The falling distance will be no more than 2 centimeters. Further improvement should be possible by performing atom chip-assisted BEC preparation as well as the interaction-suppressing mechanism~\cite{PhysRevLett.117.203003}. Gravity measurements with sub-${\mu}$Gal accuracies in miniaturized, robust devices are sure to come in the future.

\section{Conclusion}

In summary, we review our recent experimental developments on the performance of interferometer with ultra-cold atoms. First, we demonstrated a method for effective preparation of a BEC in different bands of an optical lattice within a few tens of microseconds, reducing the loading time by up to three orders of magnitude as compared to adiabatic loading. Along with this shortcut method, a Ramsey interferometer with band echo technique is employed to atoms within an OL, enormously extend the coherence time by one order of magnitude. Efforts to boost the resolution with multimode scheme is made as well. Application of a noise-resilient multi-component interferometric scheme shows that increasing the number of paths could sharpen the peaks in the time-domain interference fringes, which leads to a resolution nearly twice compared with that of a conventional double-path two-mode interferometer. We can somehow boost the momentum resolution meanwhile. The patterns in the momentum space have got an interval far less than the double recoil momentum, where the narrowest one is given as $0.6\;\hbar k_L$. However, these advancements are inseparable with our endeavor to optimize data analysis based on the PCA. Extrinsic systematic noise for absorption imaging can be reduced efficiently.
A scheme for potential compact gravimeter with ultra-cold atoms has been proposed. We believe it will tremendously shrink the size of a practical on-site instrument, promoting another widely used quantum-based technique.

\section*{Acknowledgment}
This work is supported by the National Basic Research Program of China (Grant No. 2016YFA0301501), the National Natural Science Foundation of China (Grants No. 61727819, No. 11934002, No. 91736208, and No. 11920101004), and the Project funded by China Postdoctoral Science Foundation.

\bibliographystyle{unsrt}
\bibliography{citelist}

\begin{thebibliography}{100}

\bibitem{XuRN32}
Victoria Xu, Matt Jaffe, Cristian~D. Panda, Sofus~L. Kristensen, Logan~W.
  Clark, and Holger M\"uller.
\newblock Probing gravity by holding atoms for 20 seconds.
\newblock {\em Science}, 366(6466):745, 2019.

\bibitem{CooperRN203}
N.~R. Cooper, J.~Dalibard, and I.~B. Spielman.
\newblock Topological bands for ultracold atoms.
\newblock {\em Rev. Mod. Phys.}, 91(1):015005, 2019.

\bibitem{RN114}
Linxiao Niu, Shengjie Jin, Xuzong Chen, Xiaopeng Li, and Xiaoji Zhou.
\newblock Observation of a dynamical sliding phase superfluid with $p$-band
  bosons.
\newblock {\em Phys. Rev. Lett.}, 121(26):265301, 2018.

\bibitem{MazurenkoRN164}
Anton Mazurenko, Christie~S. Chiu, Geoffrey Ji, Maxwell~F. Parsons, M\'arton
  Kan\'asz-Nagy, Richard Schmidt, Fabian Grusdt, Eugene Demler, Daniel Greif,
  and Markus Greiner.
\newblock A cold-atom fermi-hubbard antiferromagnet.
\newblock {\em Nature}, 545(7655):462, 2017.

\bibitem{1dalfovo1999theory}
Franco Dalfovo, Stefano Giorgini, Lev~P Pitaevskii, and Sandro Stringari.
\newblock Theory of bose-einstein condensation in trapped gases.
\newblock {\em Rev. Mod. Phys.}, 71(3):463, 1999.

\bibitem{BlochRN78}
Immanuel Bloch, Jean Dalibard, and Wilhelm Zwerger.
\newblock Many-body physics with ultracold gases.
\newblock {\em Rev. Mod. Phys.}, 80(3):885, 2008.

\bibitem{2005.00368}
Stuart~S. Szigeti, Samuel~P. Nolan, John~D. Close, and Simon~A. Haine.
\newblock High-precision quantum-enhanced gravimetry with a bose-einstein
  condensate.
\newblock {\em Phys. Rev. Lett.}, 125:100402, Sep 2020.

\bibitem{Hardman2016}
K.~S. Hardman, P.~J. Everitt, G.~D. McDonald, P.~Manju, P.~B. Wigley, M.~A.
  Sooriyabandara, C.~C.~N. Kuhn, J.~E. Debs, J.~D. Close, and N.~P. Robins.
\newblock Simultaneous precision gravimetry and magnetic gradiometry with a
  bose-einstein condensate: A high precision, quantum sensor.
\newblock {\em Phys. Rev. Lett.}, 117(13):138501, 2016.

\bibitem{Hardman2014}
Kyle~S. Hardman, Carlos C.~N. Kuhn, Gordon~D. McDonald, John~E. Debs, Shayne
  Bennetts, John~D. Close, and Nicholas~P. Robins.
\newblock Role of source coherence in atom interferometry.
\newblock {\em Phys. Rev. A}, 89(2):023626, 2014.

\bibitem{Ye2007}
Jun. Ye, Sebastian Blatt, Martin~M. Boyd, Seth~M. Foreman, Eric~R. Hudson,
  Tetsuya Ido, Benjamin Lev, Andrew~D. Ludlow, Brian~C. Sawyer, Benjamin Stuhl,
  and Tanya Zelinsky.
\newblock Precision measurement based on ultracold atoms and cold molecules.
\newblock {\em Int. J. Mod. Phys. D}, 16(12b):2481, 2007.

\bibitem{RN257}
T.~Berrada, S.~van Frank, R.~B\"ucker, T.~Schumm, J.~F. Schaff, and
  J.~Schmiedmayer.
\newblock Integrated mach-zehnder interferometer for bose-einstein condensates.
\newblock {\em Nat. Commun.}, 4(1):2077, 2013.

\bibitem{SchrepplerRN228}
Sydney Schreppler, Nicolas Spethmann, Nathan Brahms, Thierry Botter, Maryrose
  Barrios, and Dan~M. Stamper-Kurn.
\newblock Optically measuring force near the standard quantum limit.
\newblock {\em Science}, 344(6191):1486, 2014.

\bibitem{Xiong_2013}
Wei Xiong, Xiaoji Zhou, Xuguang Yue, Xuzong Chen, Biao Wu, and Hongwei Xiong.
\newblock Critical correlations in an ultra-cold bose gas revealed by means of
  a temporal talbot-lau interferometer.
\newblock {\em Laser Phys. Lett.}, 10(12):125502, nov 2013.

\bibitem{ColloquiumRN80}
Andrei Derevianko and Hidetoshi Katori.
\newblock Colloquium: Physics of optical lattice clocks.
\newblock {\em Rev. Mod. Phys.}, 83(2):331, 2011.

\bibitem{Campbell2017}
S.~L. Campbell, R.~B. Hutson, G.~E. Marti, A.~Goban, N.~Darkwah~Oppong, R.~L.
  McNally, L.~Sonderhouse, J.~M. Robinson, W.~Zhang, B.~J. Bloom, and J.~Ye.
\newblock A fermi-degenerate three-dimensional optical lattice clock.
\newblock {\em Science}, 358(6359):90, 2017.

\bibitem{PhysRevA.81.012115}
Xiaoji Zhou, Xia Xu, Xuzong Chen, and Jingbiao Chen.
\newblock Magic wavelengths for terahertz clock transitions.
\newblock {\em Phys. Rev. A}, 81:012115, Jan 2010.

\bibitem{PhysRevLett.124.120403}
E.~R. Moan, R.~A. Horne, T.~Arpornthip, Z.~Luo, A.~J. Fallon, S.~J. Berl, and
  C.~A. Sackett.
\newblock Quantum rotation sensing with dual sagnac interferometers in an
  atom-optical waveguide.
\newblock {\em Phys. Rev. Lett.}, 124:120403, Mar 2020.

\bibitem{PhysRevA.80.063604}
A.~Gauguet, B.~Canuel, T.~L\'ev\`eque, W.~Chaibi, and A.~Landragin.
\newblock Characterization and limits of a cold-atom sagnac interferometer.
\newblock {\em Phys. Rev. A}, 80:063604, Dec 2009.

\bibitem{LeRN260}
J.~Le~Gou\"et, T.~E. Mehlst\"aubler, J.~Kim, S.~Merlet, A.~Clairon,
  A.~Landragin, and F.~Pereira Dos~Santos.
\newblock Limits to the sensitivity of a low noise compact atomic gravimeter.
\newblock {\em Appl. Phys. B}, 92(2):133, 2008.

\bibitem{SchmidtRN261}
M.~Schmidt, A.~Senger, M.~Hauth, C.~Freier, V.~Schkolnik, and A.~Peters.
\newblock A mobile high-precision absolute gravimeter based on atom
  interferometry.
\newblock {\em Gyroscopy Navig.}, 2(3):170, 2011.

\bibitem{PhysRevLett.107.133001}
J.~K. Stockton, K.~Takase, and M.~A. Kasevich.
\newblock Absolute geodetic rotation measurement using atom interferometry.
\newblock {\em Phys. Rev. Lett.}, 107:133001, Sep 2011.

\bibitem{TackmannRN263}
G.~Tackmann, P.~Berg, C.~Schubert, S.~Abend, M.~Gilowski, W.~Ertmer, and E.~M.
  Rasel.
\newblock Self-alignment of a compact large-area atomic sagnac interferometer.
\newblock {\em New J. Phys.}, 14(1):015002, 2012.

\bibitem{AltinRN262}
P.~A. Altin, M.~T. Johnsson, V.~Negnevitsky, G.~R. Dennis, R.~P. Anderson,
  J.~E. Debs, S.~S. Szigeti, K.~S. Hardman, S.~Bennetts, G.~D. McDonald, L.~D.
  Turner, J.~D. Close, and N.~P. Robins.
\newblock Precision atomic gravimeter based on bragg diffraction.
\newblock {\em New J. Phys.}, 15(2):023009, 2013.

\bibitem{PhysRevLett.95.093202}
I.~Carusotto, L.~Pitaevskii, S.~Stringari, G.~Modugno, and M.~Inguscio.
\newblock Sensitive measurement of forces at the micron scale using bloch
  oscillations of ultracold atoms.
\newblock {\em Phys. Rev. Lett.}, 95:093202, Aug 2005.

\bibitem{7.4weitz1996multiple}
M~Weitz, T~Heupel, and TW~H{\"a}nsch.
\newblock Multiple beam atomic interferometer.
\newblock {\em Phys. Rev. Lett.}, 77(12):2356, 1996.

\bibitem{10petrovic2013multi}
Jovana Petrovic, Ivan Herrera, Pietro Lombardi, Florian Schaefer, and
  Francesco~S Cataliotti.
\newblock A multi-state interferometer on an atom chip.
\newblock {\em New J. Phys.}, 15(4):043002, 2013.

\bibitem{Hu2018}
Dong Hu, Linxiao Niu, Shengjie Jin, Xuzong Chen, Guangjiong Dong, J\"org
  Schmiedmayer, and Xiaoji Zhou.
\newblock Ramsey interferometry with trapped motional quantum states.
\newblock {\em Commun. Phys.}, 1(1):29, 2018.

\bibitem{Zhou2018}
Xiaoji Zhou, Shengjie Jin, and J\"org Schmiedmayer.
\newblock Shortcut loading a bose-einstein condensate into an optical lattice.
\newblock {\em New J. Phys.}, 20(5):055005, 2018.

\bibitem{PhysRevA.83.051608}
Bo~Lu, Thibault Vogt, Xinxing Liu, Xu~Xu, Xiaoji Zhou, and Xuzong Chen.
\newblock Cooperative scattering measurement of coherence in a spatially
  modulated bose gas.
\newblock {\em Phys. Rev. A}, 83:051608, May 2011.

\bibitem{Wang13}
Zhongkai Wang, Linxiao Niu, Peng Zhang, Mingxuan Wen, Zhen Fang, Xuzong Chen,
  and Xiaoji Zhou.
\newblock Asymmetric superradiant scattering and abnormal mode amplification
  induced by atomic density distortion.
\newblock {\em Opt. Express}, 21(12):14377, Jun 2013.

\bibitem{PhysRevA.83.053603}
Thibault Vogt, Bo~Lu, XinXing Liu, Xu~Xu, Xiaoji Zhou, and Xuzong Chen.
\newblock Mode competition in superradiant scattering of matter waves.
\newblock {\em Phys. Rev. A}, 83:053603, May 2011.

\bibitem{PhysRevA.81.013615}
Xiaoji Zhou, Fan Yang, Xuguang Yue, T.~Vogt, and Xuzong Chen.
\newblock Imprinting light phase on matter-wave gratings in superradiance
  scattering.
\newblock {\em Phys. Rev. A}, 81:013615, Jan 2010.

\bibitem{Zhou10}
Xiaoji Zhou, Xu~Xu, Lan Yin, W.~M. Liu, and Xuzong Chen.
\newblock Detecting quantum coherence of bose gases in optical lattices by
  scattering light intensity in cavity.
\newblock {\em Opt. Express}, 18(15):15664, Jul 2010.

\bibitem{PhysRevA.80.063608}
Xiaoji Zhou, Jiageng Fu, and Xuzong Chen.
\newblock High-order momentum modes by resonant superradiant scattering.
\newblock {\em Phys. Rev. A}, 80:063608, Dec 2009.

\bibitem{PhysRevA.79.033605}
Xu~Xu, Xiaoji Zhou, and Xuzong Chen.
\newblock Spectroscopy of superradiant scattering from an array of
  bose-einstein condensates.
\newblock {\em Phys. Rev. A}, 79:033605, Mar 2009.

\bibitem{LI20084750}
Juntao Li, Xiaoji Zhou, Fan Yang, and Xuzong Chen.
\newblock Superradiant rayleigh scattering from a bose-einstein condensate with
  the incident laser along the long axis.
\newblock {\em Phys. Lett. A}, 372(26):4750, 2008.

\bibitem{PhysRevA.78.052107}
Rui Guo, Xiaoji Zhou, and Xuzong Chen.
\newblock Enhancement of motional entanglement of cold atoms by pairwise
  scattering of photons.
\newblock {\em Phys. Rev. A}, 78:052107, Nov 2008.

\bibitem{PhysRevA.78.043611}
Fan Yang, Xiaoji Zhou, Juntao Li, Yuankai Chen, Lin Xia, and Xuzong Chen.
\newblock Resonant sequential scattering in two-frequency-pumping superradiance
  from a bose-einstein condensate.
\newblock {\em Phys. Rev. A}, 78:043611, Oct 2008.

\bibitem{PhysRevA.83.033620}
Bo~Lu, Xiaoji Zhou, Thibault Vogt, Zhen Fang, and Xuzong Chen.
\newblock Laser driving of superradiant scattering from a bose-einstein
  condensate at variable incidence angle.
\newblock {\em Phys. Rev. A}, 83:033620, Mar 2011.

\bibitem{FF}
Shumpei Masuda, Katsuhiro Nakamura, and Adolfo del Campo.
\newblock High-fidelity rapid ground-state loading of an ultracold gas into an
  optical lattice.
\newblock {\em Phys. Rev. Lett.}, 113(6):063003, 2014.

\bibitem{Chen}
Xi~Chen, A.~Ruschhaupt, S.~Schmidt, A.~del Campo, D.~Gu\'ery-Odelin, and J.~G.
  Muga.
\newblock Fast optimal frictionless atom cooling in harmonic traps: Shortcut to
  adiabaticity.
\newblock {\em Phys. Rev. Lett.}, 104(6):063002, 2010.

\bibitem{Liu}
Xinxing Liu, Xiaoji Zhou, Wei Xiong, Thibault Vogt, and Xuzong Chen.
\newblock Rapid nonadiabatic loading in an optical lattice.
\newblock {\em Phys. Rev. A}, 83(6):063402, 2011.

\bibitem{Zhai}
Yueyang Zhai, Xuguang Yue, Yanjiang Wu, Xuzong Chen, Peng Zhang, and Xiaoji
  Zhou.
\newblock Effective preparation and collisional decay of atomic condensates in
  excited bands of an optical lattice.
\newblock {\em Phys. Rev. A}, 87(6):063638, 2013.

\bibitem{PhysRevA.88.013603}
Xuguang Yue, Yueyang Zhai, Zhongkai Wang, Hongwei Xiong, Xuzong Chen, and
  Xiaoji Zhou.
\newblock Observation of diffraction phases in matter-wave scattering.
\newblock {\em Phys. Rev. A}, 88:013603, Jul 2013.

\bibitem{PhysRevA.83.063604}
Xinxing Liu, Xiaoji Zhou, Wei Zhang, Thibault Vogt, Bo~Lu, Xuguang Yue, and
  Xuzong Chen.
\newblock Exploring multiband excitations of interacting bose gases in a
  one-dimensional optical lattice by coherent scattering.
\newblock {\em Phys. Rev. A}, 83:063604, Jun 2011.

\bibitem{PhysRevA.88.053629}
Yueyang Zhai, Peng Zhang, Xuzong Chen, Guangjiong Dong, and Xiaoji Zhou.
\newblock Bragg diffraction of a matter wave driven by a pulsed nonuniform
  magnetic field.
\newblock {\em Phys. Rev. A}, 88:053629, Nov 2013.

\bibitem{RN107}
Wei Xiong, Xuguang Yue, Zhongkai Wang, Xiaoji Zhou, and Xuzong Chen.
\newblock Manipulating the momentum state of a condensate by sequences of
  standing-wave pulses.
\newblock {\em Phys. Rev. A}, 84(4):043616, 2011.

\bibitem{RN111}
Zhongkai Wang, Baoguo Yang, Dong Hu, Xuzong Chen, Hongwei Xiong, Biao Wu, and
  Xiaoji Zhou.
\newblock Observation of quantum dynamical oscillations of ultracold atoms in
  the $f$ and $d$ bands of an optical lattice.
\newblock {\em Phys. Rev. A}, 94(3):033624, 2016.

\bibitem{RN112}
Baoguo Yang, Shengjie Jin, Xiangyu Dong, Zhe Liu, Lan Yin, and Xiaoji Zhou.
\newblock Atomic momentum patterns with narrower intervals.
\newblock {\em Phys. Rev. A}, 94(4):043607, 2016.

\bibitem{NiuOE}
Linxiao Niu, Dong Hu, Shengjie Jin, Xiangyu Dong, Xuzong Chen, and Xiaoji Zhou.
\newblock Excitation of atoms in an optical lattice driven by polychromatic
  amplitude modulation.
\newblock {\em Opt. Express}, 23(8):10064, 2015.

\bibitem{HuPRA}
Dong Hu, Linxiao Niu, Baoguo Yang, Xuzong Chen, Biao Wu, Hongwei Xiong, and
  Xiaoji Zhou.
\newblock Long-time nonlinear dynamical evolution for $p$-band ultracold atoms
  in an optical lattice.
\newblock {\em Phys. Rev. A}, 92(4):043614, 2015.

\bibitem{GuoRN265}
Xinxin Guo, Wenjun Zhang, Zhihan Li, Hongmian Shui, Xuzong Chen, and Xiaoji
  Zhou.
\newblock Asymmetric population of momentum distribution by quasi-periodically
  driving a triangular optical lattice.
\newblock {\em Opt. Express}, 27(20):27786, 2019.

\bibitem{RN231}
Peng-Ju Tang, Peng Peng, Xiang-Yu Dong, Xu-Zong Chen, and Xiao-Ji Zhou.
\newblock Implementation of full spin-state interferometer.
\newblock {\em Chinese Phys. Lett.}, 36(5):050301, 2019.

\bibitem{RN113}
Linxiao Niu, Pengju Tang, Baoguo Yang, Xuzong Chen, Biao Wu, and Xiaoji Zhou.
\newblock Observation of quantum equilibration in dilute bose gases.
\newblock {\em Phys. Rev. A}, 94(6):063603, 2016.

\bibitem{PhysRevLett.102.100401}
Luca Pezz\'e and Augusto Smerzi.
\newblock Entanglement, nonlinear dynamics, and the heisenberg limit.
\newblock {\em Phys. Rev. Lett.}, 102:100401, Mar 2009.

\bibitem{RN256}
B.~L\"ucke, M.~Scherer, J.~Kruse, L.~Pezz\'e, F.~Deuretzbacher, P.~Hyllus,
  O.~Topic, J.~Peise, W.~Ertmer, J.~Arlt, L.~Santos, A.~Smerzi, and C.~Klempt.
\newblock Twin matter waves for interferometry beyond the classical limit.
\newblock {\em Science}, 334(6057):773, 2011.

\bibitem{PhysRevA.64.015602}
Wei-Dong Li, X.~J. Zhou, Y.~Q. Wang, J.~Q. Liang, and Wu-Ming Liu.
\newblock Time evolution of the relative phase in two-component bose-einstein
  condensates with a coupling drive.
\newblock {\em Phys. Rev. A}, 64:015602, Jun 2001.

\bibitem{13Simsarian2000}
J.~E. Simsarian, J.~Denschlag, Mark Edwards, Charles~W. Clark, L.~Deng, E.~W.
  Hagley, K.~Helmerson, S.~L. Rolston, and W.~D. Phillips.
\newblock Imaging the phase of an evolving bose-einstein condensate wave
  function.
\newblock {\em Phys. Rev. Lett.}, 85:2040, Sep 2000.

\bibitem{13machluf2013coherent}
Shimon Machluf, Yonathan Japha, and Ron Folman.
\newblock Coherent stern-gerlach momentum splitting on an atom chip.
\newblock {\em Nat. Commun.}, 4:2424, 2013.

\bibitem{12margalit2015self}
Yair Margalit, Zhifan Zhou, Shimon Machluf, Daniel Rohrlich, Yonathan Japha,
  and Ron Folman.
\newblock A self-interfering clock as a "which path" witness.
\newblock {\em Science}, 349(6253):1205, 2015.

\bibitem{Tang2019a}
Pengju Tang, Peng Peng, Zhihan Li, Xuzong Chen, Xiaopeng Li, and Xiaoji Zhou.
\newblock Parallel multicomponent interferometer with a spinor bose-einstein
  condensate.
\newblock {\em Phys. Rev. A}, 100:013618, Jul 2019.

\bibitem{29PRLinterferency}
Meng Han, Peipei Ge, Yun Shao, Qihuang Gong, and Yunquan Liu.
\newblock Attoclock photoelectron interferometry with two-color corotating
  circular fields to probe the phase and the amplitude of emitting wave
  packets.
\newblock {\em Phys. Rev. Lett.}, 120:073202, Feb 2018.

\bibitem{30RevModPhys}
Alexander~D. Cronin, J\"org Schmiedmayer, and David~E. Pritchard.
\newblock Optics and interferometry with atoms and molecules.
\newblock {\em Rev. Mod. Phys.}, 81:1051, Jul 2009.

\bibitem{31Andrews637}
M.~R. Andrews, C.~G. Townsend, H.-J. Miesner, D.~S. Durfee, D.~M. Kurn, and
  W.~Ketterle.
\newblock Observation of interference between two bose condensates.
\newblock {\em Science}, 275(5300):637, 1997.

\bibitem{21Xiuquan2006}
Xiuquan Ma, Lin Xia, Fang Yang, Xiaoji Zhou, Yiqiu Wang, Hong Guo, and Xuzong
  Chen.
\newblock Population oscillation of the multicomponent spinor bose-einstein
  condensate induced by nonadiabatic transitions.
\newblock {\em Phys. Rev. A}, 73:013624, Jan 2006.

\bibitem{46Majorana}
E.~Majorana.
\newblock Atomi orientati in campo magnetico variabile.
\newblock {\em Nuovo Cimento}, 9:43, 1932.

\bibitem{23XiaLin2008}
Lin Xia, Xu~Xu, Rui Guo, Fan Yang, Wei Xiong, Juntao Li, Qianli Ma, Xiaoji
  Zhou, Hong Guo, and Xuzong Chen.
\newblock Manipulation of the quantum state by the majorana transition in
  spinor bose-einstein condensates.
\newblock {\em Phys. Rev. A}, 77:043622, Apr 2008.

\bibitem{PhysRevA.101.013612}
Pengju Tang, Xiangyu Dong, Wenjun Zhang, Yunhong Li, Xuzong Chen, and Xiaoji
  Zhou.
\newblock Implementation of a double-path multimode interferometer using a
  spinor bose-einstein condensate.
\newblock {\em Phys. Rev. A}, 101:013612, Jan 2020.

\bibitem{7.3three-path}
Gregor Weihs, Michael Reck, Harald Weinfurter, and Anton Zeilinger.
\newblock All-fiber three-path mach--zehnder interferometer.
\newblock {\em Opt. Lett.}, 21(4):302, Feb 1996.

\bibitem{7.2multiphoton}
M.~W. Mitchell, J.~S. Lundeen, and A.~M. Steinberg.
\newblock Super-resolving phase measurements with a multiphoton entangled
  state.
\newblock {\em Nature}, 429:161, May 2004.

\bibitem{7.5PhysRevA.87.033607}
J.~Chwede\'{n}czuk, F.~Piazza, and A.~Smerzi.
\newblock Multipath interferometer with ultracold atoms trapped in an optical
  lattice.
\newblock {\em Phys. Rev. A}, 87:033607, Mar 2013.

\bibitem{8paul2017measuring}
Tania Paul and Tabish Qureshi.
\newblock Measuring quantum coherence in multislit interference.
\newblock {\em Phys. Rev. A}, 95(4):042110, 2017.

\bibitem{3.4Pikovski2017clock}
Igor Pikovski, Magdalena Zych, Fabio Costa, and {\v{C}}aslav Brukner.
\newblock Time dilation in quantum systems and decoherence.
\newblock {\em New J. Phys.}, 19(2):025011, 2017.

\bibitem{7fort2001spatial}
C~Fort, P~Maddaloni, F~Minardi, M~Modugno, and M~Inguscio.
\newblock Spatial interference of coherent atomic waves by manipulation of the
  internal quantum state.
\newblock {\em Opt. Lett.}, 26(14):1039, 2001.

\bibitem{14PhysRevA.67.053803}
Jonas S\"oderholm, Gunnar Bj\"ork, Bj\"orn Hessmo, and Shuichiro Inoue.
\newblock Quantum limits on phase-shift detection using multimode
  interferometers.
\newblock {\em Phys. Rev. A}, 67:053803, May 2003.

\bibitem{11PhysRevA2017multimode}
J.~Chwede\'{n}czuk.
\newblock Quantum interferometry in multimode systems.
\newblock {\em Phys. Rev. A}, 96:032320, Sep 2017.

\bibitem{Segal}
Stephen~R. Segal, Quentin Diot, Eric~A. Cornell, Alex~A. Zozulya, and Dana~Z.
  Anderson.
\newblock Revealing buried information: Statistical processing techniques for
  ultracold-gas image analysis.
\newblock {\em Phys. Rev. A}, 81:053601, May 2010.

\bibitem{Chiow-11}
Sheng-wey Chiow, Tim Kovachy, Hui-Chun Chien, and Mark~A. Kasevich.
\newblock $102\ensuremath{\hbar}k$ large area atom interferometers.
\newblock {\em Phys. Rev. Lett.}, 107:130403, Sep 2011.

\bibitem{Nlx2018}
Linxiao Niu, Xinxin Guo, Yuan Zhan, Xuzong Chen, Wuming Liu, and Xiaoji Zhou.
\newblock Optimized fringe removal algorithm for absorption images.
\newblock {\em Appl. Phys. Lett.}, 113(14):144103, 2018.

\bibitem{Jin2019}
Shengjie Jin, Xinxin Guo, Peng Peng, Xuzong Chen, Xiaopeng Li, and Xiaoji Zhou.
\newblock Finite temperature phase transition in a cross-dimensional triangular
  lattice.
\newblock {\em New J. Phys.}, 21(7):073015, 2019.

\bibitem{Becker2010}
C.~Becker, P.~Soltan-Panahi, J.~Kronj\"ager, S.~D\"orscher, K.~Bongs, and
  K.~Sengstock.
\newblock Ultracold quantum gases in triangular optical lattices.
\newblock {\em New J. Phys.}, 12(6):065025, 2010.

\bibitem{PhysRevA.55.R3987}
M.~Gatzke, G.~Birkl, P.~S. Jessen, A.~Kastberg, S.~L. Rolston, and W.~D.
  Phillips.
\newblock Temperature and localization of atoms in three-dimensional optical
  lattices.
\newblock {\em Phys. Rev. A}, 55:R3987, Jun 1997.

\bibitem{LEVINE}
Martin~D Levine.
\newblock Feature extraction: A survey.
\newblock {\em Proc. IEEE}, 57(8):1391, 1969.

\bibitem{PCANJP1}
Romain Dubessy, Camilla De~Rossi, Thomas Badr, Laurent Longchambon, and Helene
  Perrin.
\newblock Imaging the collective excitations of an ultracold gas using
  statistical correlations.
\newblock {\em New J. Phys.}, 16:122001, Dec 2014.

\bibitem{PCANJP2}
Andrea Alberti, Carsten Robens, Wolfgang Alt, Stefan Brakhane, Michal Karski,
  Rene Reimann, Artur Widera, and Dieter Meschede.
\newblock Super-resolution microscopy of single atoms in optical lattices.
\newblock {\em New J. Phys.}, 18:053010, May 2016.

\bibitem{PCAPRA}
C.~F. Ockeloen, A.~F. Tauschinsky, R.~J.~C. Spreeuw, and S.~Whitlock.
\newblock Detection of small atom numbers through image processing.
\newblock {\em Phys. Rev. A}, 82:061606, Dec 2010.

\bibitem{Caoshuyang}
Shuyang Cao, Pengju Tang, Xinxin Guo, Xuzong Chen, Wei Zhang, and Xiaoji Zhou.
\newblock Extraction and identification of noise patterns for ultracold atoms
  in an optical lattice.
\newblock {\em Opt. Express}, 27(9):12710, 2019.

\bibitem{macroscopic}
Oliver Penrose and Lars Onsager.
\newblock Bose-einstein condensation and liquid helium.
\newblock {\em Phys. Rev.}, 104:576, Nov 1956.

\bibitem{Gustavson2000}
T.~L. Gustavson, A.~Landragin, and M.~A. Kasevich.
\newblock Rotation sensing with a dual atom-interferometer sagnac gyroscope.
\newblock {\em Classical Quant. Grav.}, 17(12):2385, 2000.

\bibitem{PhysRevLett.100.031101}
Holger M\"uller, Sheng-wey Chiow, Sven Herrmann, Steven Chu, and Keng-Yeow
  Chung.
\newblock Atom-interferometry tests of the isotropy of post-newtonian gravity.
\newblock {\em Phys. Rev. Lett.}, 100:031101, Jan 2008.

\bibitem{Louchet2011}
Anne Louchet-Chauvet, Tristan Farah, Quentin Bodart, Andr\'e Clairon, Arnaud
  Landragin, S\'ebastien Merlet, and Franck Pereira~Dos Santos.
\newblock The influence of transverse motion within an atomic gravimeter.
\newblock {\em New J. Phys.}, 13(6):065025, 2011.

\bibitem{Fixler2007}
J.~B. Fixler, G.~T. Foster, J.~M. McGuirk, and M.~A. Kasevich.
\newblock Atom interferometer measurement of the newtonian constant of gravity.
\newblock {\em Science}, 315(5808):74, 2007.

\bibitem{PhysRevLett.100.050801}
G.~Lamporesi, A.~Bertoldi, L.~Cacciapuoti, M.~Prevedelli, and G.~M. Tino.
\newblock Determination of the newtonian gravitational constant using atom
  interferometry.
\newblock {\em Phys. Rev. Lett.}, 100:050801, Feb 2008.

\bibitem{PhysRevLett.111.151102}
Michael~A. Hohensee, Holger M\"uller, and R.~B. Wiringa.
\newblock Equivalence principle and bound kinetic energy.
\newblock {\em Phys. Rev. Lett.}, 111:151102, Oct 2013.

\bibitem{PhysRevLett.106.151102}
Michael~A. Hohensee, Steven Chu, Achim Peters, and Holger M\"uller.
\newblock Equivalence principle and gravitational redshift.
\newblock {\em Phys. Rev. Lett.}, 106:151102, Apr 2011.

\bibitem{PhysRevA.88.031605}
Manuel Andia, Rapha\"el Jannin, Francois Nez, Francois Biraben, Sa\"{\i}da
  Guellati-Kh\'elifa, and Pierre Clad\'e.
\newblock Compact atomic gravimeter based on a pulsed and accelerated optical
  lattice.
\newblock {\em Phys. Rev. A}, 88:031605, Sep 2013.

\bibitem{PhysRevA.55.2989}
Ekkehard Peik, Maxime Ben~Dahan, Isabelle Bouchoule, Yvan Castin, and
  Christophe Salomon.
\newblock Bloch oscillations of atoms, adiabatic rapid passage, and monokinetic
  atomic beams.
\newblock {\em Phys. Rev. A}, 55:2989, Apr 1997.

\bibitem{PhysRevLett.76.4508}
Maxime Ben~Dahan, Ekkehard Peik, Jakob Reichel, Yvan Castin, and Christophe
  Salomon.
\newblock Bloch oscillations of atoms in an optical potential.
\newblock {\em Phys. Rev. Lett.}, 76:4508, Jun 1996.

\bibitem{PhysRevLett.76.4512}
S.~R. Wilkinson, C.~F. Bharucha, K.~W. Madison, Qian Niu, and M.~G. Raizen.
\newblock Observation of atomic wannier-stark ladders in an accelerating
  optical potential.
\newblock {\em Phys. Rev. Lett.}, 76:4512, Jun 1996.

\bibitem{PhysRevLett.92.253001}
R\'emy Battesti, Pierre Clad\'e, Sa\"{\i}da Guellati-Kh\'elifa, Catherine
  Schwob, Beno\^{\i}t Gr\'emaud, Fran\c{c}ois Nez, Lucile Julien, and
  Fran\c{c}ois Biraben.
\newblock Bloch oscillations of ultracold atoms: A tool for a metrological
  determination of $h/{m}_{\mathrm{rb}}$.
\newblock {\em Phys. Rev. Lett.}, 92:253001, Jun 2004.

\bibitem{Peters2001}
A.~Peters, K.~Y. Chung, and S.~Chu.
\newblock High-precision gravity measurements using atom interferometry.
\newblock {\em Metrologia}, 38(1):25, 2001.

\bibitem{PhysRevA.74.052109}
Pierre Clad\'e, Estefania de~Mirandes, Malo Cadoret, Sa\"{\i}da
  Guellati-Kh\'elifa, Catherine Schwob, Fran\c{c}ois Nez, Lucile Julien, and
  Fran\c{c}ois Biraben.
\newblock Precise measurement of $h/{m}_{\mathrm{rb}}$ using bloch oscillations
  in a vertical optical lattice: Determination of the fine-structure constant.
\newblock {\em Phys. Rev. A}, 74:052109, Nov 2006.

\bibitem{PhysRevLett.117.203003}
S.~Abend, M.~Gebbe, M.~Gersemann, H.~Ahlers, H.~M\"untinga, E.~Giese,
  N.~Gaaloul, C.~Schubert, C.~L\"ammerzahl, W.~Ertmer, W.~P. Schleich, and
  E.~M. Rasel.
\newblock Atom-chip fountain gravimeter.
\newblock {\em Phys. Rev. Lett.}, 117:203003, Nov 2016.

\end{thebibliography}

\end{document}